\documentclass[aps,prc,showpacs,preprint,nobibnotes,superscriptaddress]{revtex4-1}

\usepackage{longtable}
\usepackage{graphicx}
\usepackage{dcolumn}
\usepackage{bm}
\usepackage{longtable}

\usepackage{amsmath}
\newcommand\vecl[1]{\overrightarrow{#1}}
\newcommand\cev[1]{\overleftarrow{#1}}

\begin{document}

\preprint{KEK-TH-1653}

\title{Probing gluon dynamics by charm and bottom mesons in nuclear medium 
 from heavy meson effective theory with $1/M$-corrections}


\author{S.~Yasui}
\email[]{yasuis@post.kek.jp}
\affiliation{KEK Theory Center, Institute of Particle and Nuclear
Studies, High Energy Accelerator Research Organization, 1-1, Oho,
Ibaraki, 305-0801, Japan}
\author{K.~Sudoh}
\affiliation{Nishogakusha University, 6-16, Sanbancho, Chiyoda,
Tokyo, 102-8336, Japan}


\date{\today}

\begin{abstract}
We consider heavy mesons with charm and bottom in nuclear medium.
We construct the effective Lagrangian with axial-vector coupling including $1/M$-corrections for the heavy meson mass $M$ by following the velocity-rearrangement invariance.
As an application, we consider heavy mesons, $\bar{\mathrm{D}}$ and $\bar{\mathrm{D}}^{\ast}$ mesons for charm and $\mathrm{B}$ and $\mathrm{B}^{\ast}$ mesons for bottom, bound in nuclear matter, and we discuss their in-medium masses 
modified by the interaction with nucleons via pion exchanges including the $1/M$-corrections.
The mass modifications are affected by the gluon dynamics in nuclear medium.
By comparison with the heavy quark effective theory,
we find that the effects of scale anomaly become suppressed in nuclear medium.
We also find that the contributions from the chromoelectric gluon are enhanced in nuclear medium, while those from the chromomagnetic gluon are reduced.
We propose to use heavy mesons as probes to research the gluon fields in nuclear medium in experimental studies. 
\end{abstract}


\maketitle

\section{Introduction}

Exotic nuclei containing hadrons as impurities are interesting, not only for studying hadron-nucleon interactions and changes of nuclear structures induced by the impurities, but also for investigating medium effects at finite baryon number density as modifications of the QCD vacuum.
In light flavor (up, down and strangeness) sector, it has been studied to explore the chiral condensates as well as the gluon condensates in nuclear medium by using light hadrons as probes \cite{Brown:1991kk,Cohen:1991nk,Cohen:1994wm,Birse:1994cz,Hatsuda:1991ez,Hatsuda:1994pi,Hayano:2008vn}.
As a natural extension from light flavors to heavy flavors, there have been also discussions about exotic nuclei containing charm and bottom hadrons \cite{Luke:1992tm,Klingl:1998sr,Song:2008bd,Morita:2010pd,Hayashigaki:2000es,Friman:2002fs,Hilger:2008jg,Hilger:2010zb,Wang:2011mj}.
Exotic nuclei with charm and bottom flavors will bring us new knowledge which is difficult to be accessed by those with light flavors.
It is expected that they are studied in future experiments in accelerator facilities.

We consider mass for the hadron and nuclear system containing a heavy quark with mass $m_{\mathrm{Q}}$.
It is given by $1/m_{\mathrm{Q}}$-expansion by following the heavy quark effective theory (HQET) \cite{Neubert:1993mb,Manohar:2000dt}.
Importantly, the coefficients in the power series of $1/m_{\mathrm{Q}}$ are related to the gluon dynamics in the heavy systems.
As well known, at leading order of the $1/m_{\mathrm{Q}}$-expansion (the heavy quark limit $m_{\mathrm{Q}} \rightarrow \infty$), the system with a heavy quark obeys the heavy quark symmetry, namely the heavy-flavor symmetry and the heavy-quark-spin symmetry.
At this order, the mass of the heavy system is simply a sum of the mass of the heavy quark and the energy from the light component, namely the light quarks and the gluons.
The contributions from the light component (light quarks and gluons) are related to the scale anomaly of the energy-momentum tensor in QCD \cite{Bigi:1994ga,Bigi:1997fj},
which is the analogue of the gluon condensate in the QCD vacuum.
In the present study, we further explore the corrections at $\mathcal{O}(1/m_{\mathrm{Q}}^1)$ in the $1/m_{\mathrm{Q}}$-expansion.
It provides us with another information about the gluon dynamics.
At $\mathcal{O}(1/m_{\mathrm{Q}}^1)$, ``chromomagnetic" gluon field
 is concerned with the mass splitting between a pair of spin partners, such as a $\bar{\mathrm{P}}=(\bar{\mathrm{Q}}\mathrm{q})_{\mathrm{spin}\,0}$ meson and a $\bar{\mathrm{P}}^{\ast}=(\bar{\mathrm{Q}}\mathrm{q})_{\mathrm{spin}\,1}$ meson with a heavy antiquark $\bar{\mathrm{Q}}$ and a light quark $\mathrm{q}$.
Furthermore, it is also known from the virial theorem that
``chromoelectric" gluon field is related to the kinetic energy of the heavy system 
with $\mathcal{O}(1/m_{\mathrm{Q}}^1)$
 \cite{Neubert:1993zc}.
Both of them are involved in the mass formula for the heavy system which is given by power series of $1/m_{\mathrm{Q}}$ (see Eq.~(\ref{eq:mass_formula}) in Sec.~\ref{sec:matrix_elements}).
Thus, we obtain the information about the gluon dynamics from the masses of heavy systems.
This is a quite general procedure so that it can be applied not only to heavy hadrons but also to exotic nuclei containing a heavy quark.

Let us focus on the charm (bottom) nuclei with anticharm ($C=-1$) (antibottom ($B=+1$)), which contain
a $\bar{\mathrm{D}}^{(\ast)}$ ($\mathrm{B}^{(\ast)}$) meson in the ground state.
We note that a $\bar{\mathrm{D}}^{(\ast)}$ ($\mathrm{B}^{(\ast)}$) meson is composed of $\bar{\mathrm{c}}\mathrm{q}$ ($\bar{\mathrm{b}}\mathrm{q}$) with a charm (bottom) antiquark $\bar{\mathrm{c}}$ ($\bar{\mathrm{b}}$) and a light quark q, and hence there is no annihilation process from light quark-antiquark pairs in nuclear medium.
Therefore, when the interaction between a $\bar{\mathrm{D}}^{(\ast)}$ ($\mathrm{B}^{(\ast)}$) meson and a nucleon $\mathrm{N}$ is  attractive sufficiently,
 a $\bar{\mathrm{D}}^{(\ast)}$ ($\mathrm{B}^{(\ast)}$) meson can 
 be bound in nuclei as a stable particle against the decays by strong interactions.
It decays by weak interactions and electromagnetic interactions only.
Theoretically, it was shown indeed that there is an attractive force (e.g. a pion exchange force) so that several bound and/or resonant states of $\bar{\mathrm{D}}^{(\ast)}\mathrm{N}$ ($\mathrm{B}^{(\ast)}\mathrm{N}$) can exist around the thresholds \cite{Yasui:2009bz,Yamaguchi:2011xb,Yamaguchi:2011qw,Carames:2012bd}.
The problem whether a $\bar{\mathrm{D}}^{(\ast)}$ ($\mathrm{B}^{(\ast)}$) meson is bound in nuclear matter 
 was discussed by several theoretical approaches, the quark-meson coupling models \cite{Tsushima:1998ru,Sibirtsev:1999js,Tsushima:2002cc}, the QCD sum rules \cite{Hilger:2008jg,Hilger:2010zb}, the mean-field methods \cite{Mishra:2003se,Mishra:2008cd,Kumar:2010gb,Kumar:2011ff}, the coupled-channel methods with contact interactions \cite{Lutz:2005vx,Tolos:2007vh,JimenezTejero:2011fc,GarciaRecio:2011xt} and the perturbative calculations by pion exchanges \cite{Yasui:2012rw}.
The study of atomic nuclei containing a $\bar{\mathrm{D}}^{(\ast)}$ ($\mathrm{B}^{(\ast)}$) meson was also performed.
There is also a discussion about the in-medium interaction between a $\bar{\mathrm{D}}^{(\ast)}$ ($\mathrm{B}^{(\ast)}$) meson and a nucleon in nuclear matter \cite{Yasui:2013xr}.
Many of them suggest that $\bar{\mathrm{D}}^{(\ast)}$ ($\mathrm{B}^{(\ast)}$) mesons can be bound stably.

The purpose in the present study is to investigate the in-medium masses of $\bar{\mathrm{D}}^{(\ast)}$ ($\mathrm{B}^{(\ast)}$) mesons in nuclear matter.
The previous studies about $\bar{\mathrm{D}}^{(\ast)}$ ($\mathrm{B}^{(\ast)}$) mesons in nuclear matter did not give all the terms with $\mathcal{O}(1/m_{\mathrm{Q}})$ in $1/m_{\mathrm{Q}}$-corrections.
However, it is necessary to  fully include the complete terms with $\mathcal{O}(1/m_{\mathrm{Q}})$ for discussions in realistic situations, which are comparable with experimental results.
Furthermore,
in the power series by $1/m_{\mathrm{Q}}$ for the in-medium masses,
we obtain the information about the modifications of
gluon dynamics as the medium effects.
For the achievement of this, the correct $1/m_{\mathrm{Q}}$-expansion for the masses of $\bar{\mathrm{D}}^{(\ast)}$ ($\mathrm{B}^{(\ast)}$) mesons in nuclear medium is needed.
We discuss the $1/M$-expansion with the heavy meson mass $M$ in the heavy meson effective theory (HMET) \cite{Luke:1992cs,Kitazawa:1993bk} (see also Ref.~\cite{Boyd:1994pa,Stewart:1998ke}).
In fact, it will turn out that the $1/M$-expansion up to $\mathcal{O}(1/M^1)$ corresponds to the $1/m_{\mathrm{Q}}$-expansion up to $\mathcal{O}(1/m_{\mathrm{Q}}^1)$ in the heavy quark effective theory.
We leave a comment that,
at finite density, a $\mathrm{D}^{(\ast)}$ ($\bar{\mathrm{B}}^{(\ast)}$) meson is unstable in nuclear medium, 
 because the absorption and annihilation processes are dynamically caused by light quark-antiquark pairs.
In those views, we regard that $\bar{\mathrm{D}}$ and $\bar{\mathrm{D}}^{\ast}$ ($\mathrm{B}$ and $\mathrm{B}^{\ast}$) mesons are  simple and unique objects to probe the gluon dynamics 
 in nuclear medium.

The article is organized as follows.
In Sec.~\ref{sec:matrix_elements}, we summarize briefly the mass formula  with the $1/m_{\mathrm{Q}}$-expansion
for heavy hadrons in the heavy quark effective theory. 
In Sec.~\ref{sec:heavy_meson_effective_theory}, we formulate the effective Lagrangian with the $1/M$-corrections for the heavy mesons interacting with a pion by the coupling through the axial-vector current.
In Sec.~\ref{sec:nuclear_matter}, we apply the heavy meson effective Lagrangian to calculate the in-medium mass of the $\bar{\mathrm{D}}^{(\ast)}$ ($\mathrm{B}^{(\ast)}$) meson in nuclear matter.
We then discuss the modifications of effects of the scale anomaly and the chromoelectric and chromomagnetic gluon fields in nuclear matter.
In Sec.~\ref{sec:discussion}, we discuss the application to baryons containing a heavy quark.
The final section is devoted to a summary and perspectives.

\section{Mass formula for heavy hadrons in heavy quark effective theory}
\label{sec:matrix_elements}

In the heavy quark effective theory \cite{Neubert:1993mb,Manohar:2000dt}, the four-momentum of the heavy quark Q with mass $m_{\mathrm{Q}}$ is separated as
\begin{eqnarray}
p^{\mu} = m_{\mathrm{Q}} v^{\mu} + k^{\mu},
\label{eq:momentum1}
\end{eqnarray}
with the four-velocity $v^{\mu}$ ($v^2=1$) and the residual four-momentum $k^{\mu}$ whose scale is much smaller than $m_{\mathrm{Q}}v^{\mu}$.
Then, we introduce the effective field for the heavy quark
\begin{eqnarray}
Q_{v}(x) = e^{im_{\mathrm{Q}} v \cdot x} \frac{1+v\hspace{-0.5em}/}{2} Q(x),
\end{eqnarray}
for the original heavy quark field $Q(x)$.
The effective Lagrangian including $\mathcal{O}(1/m_{\mathrm{Q}}^1)$
 is given by
\begin{eqnarray}
\mathcal{L}_{\mathrm{HQET}} &=& \overline{Q}_{v} v \!\cdot\! iD Q_{v} + \overline{Q}_{v} \frac{(iD_{\perp})^2}{2m_{\mathrm{Q}}} Q_{v} \nonumber \\
&& - c(\mu) g_{\mathrm{s}} \, \overline{Q}_{v} \frac{\sigma_{\alpha\beta}G^{\alpha\beta}}{4m_{\mathrm{Q}}}Q_{v} + \mathcal{O}(1/m_{\mathrm{Q}}^2),
\label{eq:HQET_Lagrangian}
\end{eqnarray}
with $D_{\perp}^{\mu} = D^{\mu} - v^{\mu} \, v \cdot D$ for the covariant derivative $D^{\mu} = \partial^{\mu} + ig_{\mathrm{s}} A^{\mu}$ with the coupling constant $g_{\mathrm{s}}$ and the gluon field $A^{\mu}=A^{a\,\mu}T^{a}$ ($a=1,\dots,8$).
In the third term, we define the gluon field tensor $G^{\alpha\beta}$ given by $[D^{\alpha},D^{\beta}] = ig_{\mathrm{s}} G^{\alpha\beta}$, and introduce the Wilson coefficient $c(\mu)$ from the matching to QCD at energy scale $\mu \simeq m_{\mathrm{Q}}$.
The heavy quark symmetry (the heavy-flavor symmetry and the heavy-quark-spin symmetry) is conserved at $\mathcal{O}(1/m_{\mathrm{Q}}^0)$, while it is not generally conserved at $\mathcal{O}(1/m_{\mathrm{Q}}^1)$.
In the terms with $\mathcal{O}(1/m_{\mathrm{Q}}^1)$ in Eq.~(\ref{eq:HQET_Lagrangian}), the first term breaks the heavy-flavor symmetry but still conserves the heavy-quark-spin symmetry, while the second term breaks both symmetries.
The QCD Lagrangian for light quarks and gluons is unchanged.

Based on the effective Lagrangian (\ref{eq:HQET_Lagrangian}), the mass of the hadron H containing a heavy quark Q is given as
\begin{eqnarray}
M_{\mathrm{H}} &=& m_{\mathrm{Q}} + \bar{\Lambda} - \frac{\lambda_{1}}{2m_{\mathrm{Q}}} + 4\vec{S}_{\mathrm{Q}} \!\cdot\! \vec{S}_{\mathrm{L}} \frac{\lambda_{2}(m_{\mathrm{Q}})}{2m_{\mathrm{Q}}} \nonumber \\
&& + \mathcal{O}(1/m_{\mathrm{Q}}^2),
\label{eq:mass_formula}
\end{eqnarray}
where we define, in the rest frame with $v_{\mathrm{r}}=(1,\vec{0}\,)$,
\begin{eqnarray}
  \frac{1}{2} \langle \mathrm{H}_{v_{\mathrm{r}}} | \mathcal{H}_{0}| \mathrm{H}_{v_{\mathrm{r}}} \rangle &=& \bar{\Lambda},
\label{eq:lambda0} \\
  \frac{1}{2} \langle \mathrm{H}_{v_{\mathrm{r}}} | \overline{Q}_{v_{\mathrm{r}}} (iD_{\perp})^2 Q_{v_{\mathrm{r}}} | \mathrm{H}_{v_{\mathrm{r}}} \rangle &=&  \lambda_{1}, 
\label{eq:chromoelectric0} \\
   \frac{1}{2} c(\mu) \langle \mathrm{H}_{v_{\mathrm{r}}} | \overline{Q}_{v_{\mathrm{r}}} g_{\mathrm{s}} \sigma_{\alpha\beta} G^{\alpha\beta} Q_{v_{\mathrm{r}}} | \mathrm{H}_{v_{\mathrm{r}}} \rangle &=& 8 \vec{S}_{\mathrm{Q}} \!\cdot\! \vec{S}_{\mathrm{L}} \lambda_{2}(m_{\mathrm{Q}}), \label{eq:chromomagnetic0}
\end{eqnarray}
with denoting the hadron state by $| \mathrm{H}_{v_{\mathrm{r}}} \rangle$.
The factor $1/2$ is multiplied due to the normalization of the wave function.
Here $\mathcal{H}_{0}$ is the Hamiltonian obtained from the leading term in $\mathcal{L}_{\mathrm{HQET}}$.
In Eq.~(\ref{eq:chromomagnetic0}), $\vec{S}_{\mathrm{Q}}$ and $\vec{S}_{\mathrm{L}}$ are operators for the spin of the heavy quark Q and the total angular momentum of the light degrees of freedom (the brown muck \cite{Neubert:1993mb,Manohar:2000dt} or the spin-complex \cite{Yasui:2013vca}), respectively.
The dependence of $m_{\mathrm{Q}}$ on $\lambda_{2}(m_{\mathrm{Q}})$ originates from the dependence of $m_{\mathrm{Q}}$ on the Wilson coefficient $c(\mu)$, because the matching with QCD is done at the energy scale $\mu\simeq m_{\mathrm{Q}}$.
For Eqs.~(\ref{eq:lambda0}), (\ref{eq:chromoelectric0}) and (\ref{eq:chromomagnetic0}), interestingly, there are alternative expressions given as
\begin{eqnarray}
 \frac{1}{2M_{\mathrm{H}}} \langle \tilde{\mathrm{H}}_{v_{\mathrm{r}}} | \frac{\beta(\alpha_{\mathrm{s}})}{4\alpha_{\mathrm{s}}} G^{2} | \tilde{\mathrm{H}}_{v_{\mathrm{r}}} \rangle &=& \bar{\Lambda}, 
\label{eq:lambda} \\
 \langle \mathrm{H}_{v_{\mathrm{r}}} | \overline{Q}_{v_{\mathrm{r}}} g_{\mathrm{s}} \vec{x} \!\cdot\! \vec{E} Q_{v_{\mathrm{r}}} | \mathrm{H}_{v_{\mathrm{r}}} \rangle &=& - \frac{\lambda_{1}}{m_{\mathrm{Q}}}, 
\label{eq:chromoelectric} \\
 \frac{1}{2} c(\mu) \langle \mathrm{H}_{v_{\mathrm{r}}} | \overline{Q}_{v_{\mathrm{r}}} g_{\mathrm{s}} \vec{\sigma} \!\cdot\! \vec{B} Q_{v_{\mathrm{r}}} | \mathrm{H}_{v_{\mathrm{r}}} \rangle &=& 8 \vec{S}_{\mathrm{Q}} \!\cdot\! \vec{S}_{\mathrm{L}} \lambda_{2}(m_{\mathrm{Q}}), \label{eq:chromomagnetic}
\end{eqnarray}
where $E^{i}=-G^{0i}$ is the chromoelectric gluon field and $B^{i}=\varepsilon^{ijk}G^{jk}$ is the chromomagnetic gluon field ($i,j,k=1,2,3$).
The first equation (\ref{eq:lambda}) originates from the scale anomaly in the trace of the energy-momentum tensor in QCD \cite{Bigi:1994ga,Bigi:1997fj}.
We introduce  the Gell-Mann--Low function $\beta(\alpha_{\mathrm{s}})=\mu \mathrm{d}\alpha_{\mathrm{s}}(\mu)/\mathrm{d}\mu$.
We ignore the finite current mass of light quarks for simplicity.
In this case, the light quark fields do not appear in the trace of the energy-momentum tensor.
In Eq.~(\ref{eq:lambda}), we use the state $| \tilde{\mathrm{H}}_{v_{\mathrm{r}}} \rangle$, which normalization factor is consistent with the one used in Refs.~\cite{Bigi:1994ga,Bigi:1997fj}, instead of the state $| \mathrm{H}_{v_{\mathrm{r}}} \rangle$.
The choice of the normalization factor does not affect our conclusion.
We may note that Eq.~(\ref{eq:lambda}) is an analogue of the gluon condensate in the QCD vacuum.
Instead of the real vacuum as the ground state of QCD, however, we consider the state containing a heavy quark in the present discussion. 
The second equation (\ref{eq:chromoelectric}) is derived from the virial theorem as shown in Ref.~\cite{Neubert:1993zc} (see also Ref.~\cite{Bigi:1997fj}).
Here $\vec{x}$ denotes the position of the center-of-mass of the system, which should coincide with the position of the heavy quark in the heavy quark limit.
The third equation (\ref{eq:chromomagnetic}) is straightforwardly obtained, because $\sigma_{\alpha\beta}$ contains the Pauli matrices $\vec{\sigma}$ for the heavy quark spin.
Thus, we find that the matrix elements $\bar{\Lambda}$, $\lambda_{1}$ and $\lambda_{2}(m_{\mathrm{Q}})$ are related to the scale anomaly and the chromoelectric and chromomagnetic gluon fields around the heavy quark, respectively.
This provides us with an interesting view to study the gluon dynamics.
We may use heavy hadrons to probe the gluon field (the chromoelectric and chromomagnetic fields) around the heavy quarks in vacuum.

The mass formula in Eq.~(\ref{eq:mass_formula}) does not depend on the specific structure of heavy hadrons, when the matrix elements $\bar{\Lambda}$, $\lambda_{1}$ and $\lambda_{2}(m_{\mathrm{Q}})$ are appropriately given.
It can be applied to the heavy hadrons in finite temperature and/or baryon number density, as far as the energy scale of the temperature and the baryon number density is much smaller than the heavy quark mass.
We can even consider possible bound/resonant states containing a heavy quark in the deconfinement phase.
When the heavy hadron is embedded in the medium with temperature $T$ and baryon number density $\rho$, the mass formula is given as
\begin{eqnarray}
 M_{\mathrm{H}}(T,\rho) &=& m_{\mathrm{Q}} + \bar{\Lambda}(T,\rho) - \frac{\lambda_{1}(T,\rho)}{2m_{\mathrm{Q}}} 
 + 4\vec{S}_{\mathrm{Q}} \!\cdot\! \vec{S}_{\mathrm{L}} \frac{\lambda_{2}(T,\rho;m_{\mathrm{Q}})}{2m_{\mathrm{Q}}} \nonumber \\
&&  + \mathcal{O}(1/m_{\mathrm{Q}}^2),
\label{eq:mass_formula_density}
\end{eqnarray}
where the matrix elements are defined by
\begin{eqnarray}
&& \frac{1}{2M_{\mathrm{H}}} \langle \tilde{\mathrm{H}}_{v_{\mathrm{r}}}(T,\rho) | \frac{\beta(\alpha_{\mathrm{s}})}{4\alpha_{\mathrm{s}}} G^{2} | \tilde{\mathrm{H}}_{v_{\mathrm{r}}}(T,\rho) \rangle = \bar{\Lambda}(T,\rho),
\label{eq:lambda_density} \\
&& \langle \mathrm{H}_{v_{\mathrm{r}}}(T,\rho) | \overline{Q}_{v_{\mathrm{r}}} g_{\mathrm{s}} \vec{x} \!\cdot\! \vec{E} Q_{v_{\mathrm{r}}} | \mathrm{H}_{v_{\mathrm{r}}}(T,\rho) \rangle = -\frac{\lambda_{1}(T,\rho)}{m_{\mathrm{Q}}}, 
\label{eq:chromoelectric_density} \\
&& \frac{1}{2} c(\mu) \langle \mathrm{H}_{v_{\mathrm{r}}}(T,\rho) | \overline{Q}_{v_{\mathrm{r}}} g_{\mathrm{s}} \vec{\sigma} \!\cdot\! \vec{B} Q_{v_{\mathrm{r}}} | \mathrm{H}_{v_{\mathrm{r}}}(T,\rho) \rangle \nonumber \\
&&\hspace{10.5em} = 8 \vec{S}_{\mathrm{Q}} \!\cdot\! \vec{S}_{\mathrm{L}} \lambda_{2}(T,\rho;m_{\mathrm{Q}}).
\label{eq:chromomagnetic_density}
\end{eqnarray}
for the hadron state $| \mathrm{H}_{v_{\mathrm{r}}}(T,\rho) \rangle$ in the medium.
Thus, the matrix elements $\bar{\Lambda}(T,\rho)$, $\lambda_{1}(T,\rho)$ and $\lambda_{2}(T,\rho;m_{\mathrm{Q}})$ bring us with the information about the gluon field around the heavy quark in the medium.
When we know $\bar{\Lambda}(T,\rho)$, $\lambda_{1}(T,\rho)$ and $\lambda_{2}(T,\rho;m_{\mathrm{Q}})$ in the medium, we immediately obtain the ratios
\begin{eqnarray}
 \frac{\langle \tilde{\mathrm{H}}_{v_{\mathrm{r}}}(T,\rho) | \frac{\beta(\alpha_{\mathrm{s}})}{4\alpha_{\mathrm{s}}} G^{2} | \tilde{\mathrm{H}}_{v_{\mathrm{r}}}(T,\rho) \rangle}
 {\langle \tilde{\mathrm{H}}_{v_{\mathrm{r}}} | \frac{\beta(\alpha_{\mathrm{s}})}{4\alpha_{\mathrm{s}}} G^{2} | \tilde{\mathrm{H}}_{v_{\mathrm{r}}} \rangle} &=& \frac{\bar{\Lambda}(T,\rho)}{\bar{\Lambda}},
\label{eq:ratio_lambda} \\
 \frac{\langle \mathrm{H}_{v_{\mathrm{r}}}(T,\rho) | \overline{Q}_{v_{\mathrm{r}}} g_{\mathrm{s}} \vec{x} \!\cdot\! \vec{E} Q_{v_{\mathrm{r}}} | \mathrm{H}_{v_{\mathrm{r}}}(T,\rho) \rangle}{\langle \mathrm{H}_{v_{\mathrm{r}}} | \overline{Q}_{v_{\mathrm{r}}} g_{\mathrm{s}} \vec{x} \!\cdot\! \vec{E} Q_{v_{\mathrm{r}}} | \mathrm{H}_{v_{\mathrm{r}}} \rangle}
 &=&
 \frac{\lambda_{1}(T,\rho)}{\lambda_{1}},
\label{eq:ratio1} \\
\frac{\langle \mathrm{H}_{v_{\mathrm{r}}}(T,\rho) | \overline{Q}_{v_{\mathrm{r}}} g_{\mathrm{s}} \vec{\sigma} \!\cdot\! \vec{B} Q_{v_{\mathrm{r}}} | \mathrm{H}_{v_{\mathrm{r}}}(T,\rho) \rangle}{\langle \mathrm{H}_{v_{\mathrm{r}}} | \overline{Q}_{v_{\mathrm{r}}} g_{\mathrm{s}} \vec{\sigma} \!\cdot\! \vec{B} Q_{v_{\mathrm{r}}} | \mathrm{H}_{v_{\mathrm{r}}} \rangle}
&=&
\frac{\lambda_{2}(T,\rho;m_{\mathrm{Q}})}{\lambda_{2}(m_\mathrm{Q})}, \label{eq:ratio2}
\end{eqnarray}
in comparison with $\bar{\Lambda}$, $\lambda_{1}$ and $\lambda_{2}(m_{\mathrm{Q}})$ in vacuum.
These ratios tell us how the effects of scale anomaly and chromoelectric and chromomagnetic gluon fields are changed in the medium.
Thus, it will enable us to know the gluon dynamics in the medium in comparison with that in vacuum.
The procedure is generally as follows.
First, 
 we evaluate the matrix elements $\bar{\Lambda}$, $\lambda_{1}$ and $\lambda_{2}(m_{\mathrm{Q}})$ for heavy hadrons in vacuum from the mass formula in Eq.~(\ref{eq:mass_formula}). 
Second, we do the same procedure for $\bar{\Lambda}(T,\rho)$, $\lambda_{1}(T,\rho)$ and $\lambda_{2}(T,\rho;m_{\mathrm{Q}})$ in medium from Eq.~(\ref{eq:mass_formula_density}).
Third, by using Eqs.~(\ref{eq:ratio_lambda}), (\ref{eq:ratio1}) and (\ref{eq:ratio2}), we finally find the modification of the effects of scale anomaly and chromoelectric and chromomagnetic gluon fields
 in the medium.
We note that this procedure is model-independent, 
because the mass formulae in Eqs.~(\ref{eq:mass_formula}) and (\ref{eq:mass_formula_density}) are based directly on the HQET.
It will be applied also to nonuniform medium, few-body hadron and nuclear systems and so on, as far as the systems contain a heavy quark.
Because we concentrate on nuclear matter with zero temperature ($T=0$) in the present study,
for simplicity we introduce the notations
\begin{eqnarray}
M_{\mathrm{H}}(\rho) &=& M_{\mathrm{H}}(T=0,\rho), \\
| \mathrm{H}_{v_{\mathrm{r}}}(\rho) \rangle &=& | \mathrm{H}_{v_{\mathrm{r}}}(T=0,\rho) \rangle, \\
\bar{\Lambda}(\rho) &=& \bar{\Lambda}(T=0,\rho), \\
\lambda_{1}(\rho) &=& \lambda_{1}(T=0,\rho), \\
\lambda_{2}(\rho;m_{\mathrm{Q}}) &=& \lambda_{2}(T=0,\rho;m_{\mathrm{Q}}),
\end{eqnarray}
which will be used in the following discussions.

\section{Heavy meson effective theory with $1/M$-correction}
\label{sec:heavy_meson_effective_theory}

\subsection{Correspondence between heavy quark effective theory and heavy meson effective theory}

We consider $\bar{\mathrm{D}}$ and $\bar{\mathrm{D}}^{\ast}$ ($\mathrm{B}$ and $\mathrm{B}^{\ast}$) mesons as the simplest systems in nuclear medium.
Their quark contents are given as $\bar{\mathrm{Q}}\mathrm{q}$ with a heavy {\it antiquark} $\bar{\mathrm{Q}}$ and a light quark q, in which the latter is indeed a non-perturbative object (the spin-complex \cite{Yasui:2013vca}) with a superposition of not only light quark-antiquark pairs and gluons like $\mathrm{q}+\bar{\mathrm{q}}\mathrm{qq}+\bar{\mathrm{q}}\mathrm{q}\mathrm{g}+\dots$  but also nucleon-hole pairs around the Fermi surface.
In theoretical analysis, the most reliable way to evaluate the matrix elements in Eqs.~(\ref{eq:lambda_density}), (\ref{eq:chromoelectric_density}) and (\ref{eq:chromomagnetic_density}) will be to solve directly the non-perturbative problems in QCD, such as in lattice simulations.
However, we will discuss the in-medium effects in nuclear matter, for which the lattice simulations cannot be applied yet.
In the present paper, we consider the approach from the heavy meson effective theory by introducing the hadronic degrees of freedom.
Based on the effective Lagrangian for heavy mesons, we calculate the in-medium masses of $\bar{\mathrm{D}}^{(\ast)}$ ($\mathrm{B}^{(\ast)}$) mesons in nuclear matter.
By fitting them to the in-medium mass formula in Eq.~(\ref{eq:mass_formula_density}), we can evaluate the desired matrix elements.

To use the heavy meson effective theory has an advantage, not only for performing a practical calculation, but also for keeping a consistency with the heavy quark effective theory. 
The in-medium mass formula in  Eq.~(\ref{eq:mass_formula_density}) is given by power series of $1/m_{\mathrm{Q}}$. 
Interestingly, the contributions up to $\mathcal{O}(1/m_{\mathrm{Q}}^{1})$ from the heavy quark effective theory 
correspond to the ones up to $\mathcal{O}(1/M^{1})$ in the heavy meson effective theory, with the heavy meson mass $M=(M_{\mathrm{P}}+3M_{\mathrm{P}^{\ast}})/4$ as an averaged mass of $M_{\mathrm{P}}$ for a P (pseudoscalar) meson and $M_{\mathrm{P}^{\ast}}$ for a P$^{\ast}$ (vector) meson.
In the heavy quark limit, the term from the heavy quark effective theory coincide with that from the heavy meson effective theory, because there is no heavy quark mass in both theories.
We also confirm the correspondence at $\mathcal{O}(1/m_{\mathrm{Q}})$ and $\mathcal{O}(1/M)$ by the relation
$1/M = 1/m_{\mathrm{Q}} + \mathcal{O}(1/m_{\mathrm{Q}}^2)$,
or inversely
$1/m_{\mathrm{Q}} = 1/M + \mathcal{O}(1/M^2)$,
from the mass formula in vacuum in Eq.~(\ref{eq:mass_formula}) and also in medium in Eq.~(\ref{eq:mass_formula_density}).
The contributions at $\mathcal{O}(1/m_{\mathrm{Q}}^2)$ and $\mathcal{O}(1/M^2)$ are neglected in the present accuracy.
Therefore, we can use the $1/M$-expansion in the heavy meson effective Lagrangian up to $\mathcal{O}(1/M^1)$ with keeping the correspondence to the $1/m_{\mathrm{Q}}$-expansion in the heavy quark effective theory.
In the literature, the systematic analysis to include the $1/M$-corrections in the heavy meson effective theory was given in Ref.~\cite{Luke:1992cs,Kitazawa:1993bk}.
In the next two subsections, we give their results with adding more information.

\subsection{Velocity-rearrangement invariance}

In the heavy meson effective theory, we define the meson-field
\begin{eqnarray}
H_{v}(x) = \frac{1+v\hspace{-0.5em}/}{2} \left[ P_{v}^{\ast}(x)\hspace{-2.4em}/ \hspace{1.8em} + i\gamma_{5} P_{v}(x) \right],
\end{eqnarray}
which is a superposition of the vector field $P_{v}^{\ast\,\mu}(x)$ ($\mu=0,\dots,3$) for a P$^{\ast}$ meson and the pseudoscalar field $P_{v}(x)$ for a P meson with the four-velocity $v^{\mu}$  ($v^2=1$) \cite{Isgur:1991wq,Burdman:1992gh,Wise:1992hn,Yan:1992gz,Nowak:1992um} ( see also Refs.~\cite{Casalbuoni:1996pg,Manohar:2000dt} for reviews).
We assign $\mathrm{P}^{(\ast)}$ to stand for $\mathrm{D}^{(\ast)}$ and $\bar{\mathrm{B}}^{(\ast)}$ mesons by following the notations used in the literature.
The results for $\bar{\mathrm{D}}^{(\ast)}$ and $\mathrm{B}^{(\ast)}$ mesons are easily obtained by charge transformation.
In this formalism, in a similar way to Eq.~(\ref{eq:momentum1}), the four-momentum of the $\mathrm{P}^{(\ast)}$ meson with mass $M$ is expressed as
\begin{eqnarray}
 p^{\mu} = M v^{\mu} + k^{\mu},
\end{eqnarray}
with the four-velocity $v^{\mu}$ and the residual four-momentum $k^{\mu}$.
The residual four-momentum is a quantity much smaller than the scale of $Mv^{\mu}$.
However, such a separation is not uniquely fixed, when we allow the small change of $v^{\mu}$ and $k^{\mu}$ to take into the $1/M$-correction.
Namely, we can define a new four-velocity $w^{\mu}$ instead of $v^{\mu}$;
\begin{eqnarray}
v^{\mu}=w^{\mu}-q^{\mu}/M,
\label{eq:four_velocity_change}
\end{eqnarray}
together with the replacement of the residual momentum from $k^{\mu}$ to $k^{\mu}+q^{\mu}$, under the constraint $v\cdot q = \mathcal{O}(q^{2}/M)$.
This transformation is called the velocity-rearrangement \cite{Luke:1992cs}.
Under the velocity-rearrangement, we keep $w^2=1$ as
\begin{eqnarray}
w^2=(v+q/M)^2=1+\mathcal{O}(1/M^2),
\end{eqnarray}
with dropping the terms at $\mathcal{O}(1/M^2)$.
In accompany with the velocity-rearrangement,
the meson-fields $H_{v}(x)$ and $H_{w}(x)$ are related as
\begin{eqnarray}
H_{v}(x) &=& \left[ H_{w}(x) - \frac{1}{2M} \left[ q\hspace{-0.5em}/,H_{w}(x) \right] \right] e^{-iq\cdot x} \nonumber \\
&& + \mathcal{O}(1/M^2).
\label{eq:transform_H}
\end{eqnarray}
This relation is obtained from the Lorentz transformations from the frame with $w^{\mu}$ to the frame with $v^{\mu}$, as explicitly shown in Ref.~\cite{Kitazawa:1993bk}.
Hereafter we do not write the position $x$ in the fields for simplicity.

The velocity-rearrangement allows us an arbitrary choice of the four-velocity up to $\mathcal{O}(1/M)$. 
Accordingly, the Lagrangian should be invariant under the velocity-rearrangement,
 which is called the velocity-rearrangement invariance \cite{Luke:1992cs}.
This is a general concept which can be applied to any heavy particles in order for that the corrections by finite mass are systematically included.
We note that, in the heavy quark effective theory, the Lagrangian (\ref{eq:HQET_Lagrangian}) is invariant under the velocity rearrangement by replacing $M$ with $m_{\mathrm{Q}}$ in Eq.~(\ref{eq:four_velocity_change}) (see Ref.~\cite{Neubert:1993mb,Manohar:2000dt}).
Let us find the heavy meson effective Lagrangian which is a velocity-rearrangement invariant.
However, one may think that this may not be straightforwardly accomplished, because
 $H_{w}$ accompanies the additional term, $-\left[ q\hspace{-0.5em}/,H_{w}(x) \right]/2M$, in Eq.~(\ref{eq:transform_H}).
Then, one may think that it is not an easy task to find the Lagrangian invariant in the velocity-rearrangement.
A useful method is to introduce the ``covariant meson-field" defined as
\begin{eqnarray}
{\mathcal H}_{v} &=& H_{v} + \frac{1}{2M} \left( i\vecl{D}\hspace{-0.7em}/ H_{v} - H_{v} i\cev{D}\hspace{-0.7em}/ - 2v\!\cdot \!iD H_{v} \right) \nonumber \\
&& + \mathcal{O}(1/M^2).
\end{eqnarray}
by following the prescription in Ref.~\cite{Kitazawa:1993bk}.
Here the fact that $v^{\mu}+iD^{\mu}/M$ is a velocity-rearrangement invariant plays an important role.
We define $ i\vecl{D}\hspace{-0.7em}/ H_{v} =  i \gamma^{\mu} D_{\mu} H_{v}$ and $H_{v} i\cev{D}\hspace{-0.7em}/ = i D_{\mu} H_{v} \gamma^{\mu}$.
We confirm that ${\mathcal H}_{v}$ and ${\mathcal H}_{w}$ are related by
\begin{eqnarray}
{\mathcal H}_{v} = e^{-i q\cdot x} {\mathcal H}_{w} + \mathcal{O}(1/M^2).
\end{eqnarray}
Therefore, we have only the phase factor $e^{-i q\cdot x}$ in the transformation of the velocity-rearrangement. 
We also define the ``covariant four-velocity"
\begin{eqnarray}
{\mathcal V}^{\mu} = \frac{v^{\mu}+iD^{\mu}/M}{|v^{\mu}+iD^{\mu}/M|},
\end{eqnarray}
with the normalization condition
\begin{eqnarray}
{\mathcal V}^{\mu}{\mathcal V}_{\mu} = 1.
\end{eqnarray}
At $\mathcal{O}(1/M^1)$, ${\mathcal V}^{\mu}$ is approximated by
\begin{eqnarray}
{\mathcal V}^{\mu} = v^{\mu} + \frac{1}{M} \left( iD^{\mu} - v^{\mu} \,v \!\cdot\! iD \right) + {\mathcal O}(1/M^2),
\end{eqnarray}
which satisfies the normalization condition
\begin{eqnarray}
{\mathcal V}^{\mu}{\mathcal V}_{\mu} = 1 + {\mathcal O}(1/M^2),
\end{eqnarray}
in the desired accuracy.
When ${\mathcal V}^{\mu}$ is operated to the field $H_{v}$ in $\mathcal{H}_{v}$, we have
\begin{eqnarray}
 {\mathcal V}^{\mu} H_{v} 
&=& \left( v^{\mu} + \frac{1}{M} \left( iD^{\mu} - v^{\mu} \,v \!\cdot\! iD \right) \right) H_{v} \nonumber \\
&& + {\mathcal O}(1/M^2),
\end{eqnarray}
where the derivative $\partial^{\mu}$ in $D^{\mu}$ gives the residual momentum.
When ${\mathcal V}^{\mu}$ is operated to the conjugate field of $H_{v}$, namely $\bar{H}_{v}$, we define
\begin{eqnarray}
{\mathcal V}^{\mu} \bar{H}_{v} &=& \left( v^{\mu} - \frac{1}{M} \left( iD^{\mu} - v^{\mu} \,v \!\cdot\! iD \right) \right) \bar{H}_{v} \nonumber \\
&& + {\mathcal O}(1/M^2).
\end{eqnarray}
The covariant meson-field and four-velocity satisfy the relations
\begin{eqnarray}
\mathcal{V}\hspace{-0.55em}/ \mathcal{H}_{v}(x) &=& \mathcal{H}_{v}(x) + \mathcal{O}(1/M^2), \\
\mathcal{H}_{v}(x) \cev{\mathcal{V}}\hspace{-0.7em}/ &=& - \mathcal{H}_{v}(x) + \mathcal{O}(1/M^2),
\end{eqnarray}
as a generalization of $v\hspace{-0.5em}/H_{v}(x)=H_{v}(x)$ and $H_{v}(x)v\hspace{-0.5em}/=-H_{v}(x)$ in the original field and four-velocity.
With these setups, one can construct the Lagrangian, which is invariant under the velocity-rearrangement,  
 up to $\mathcal{O}(1/M^1)$.

\subsection{Axial currents with $1/M$-corrections}

We consider the interaction vertices for a P$^{(\ast)}$ meson and a pion through the axial-vector current coupling.
As general forms of the axial-vector currents for the heavy mesons, we may consider the forms of $\mathrm{Tr}\, \overline{\mathcal{H}}_{v} \Gamma \mathcal{H}_{v} \Gamma'$ with possible combinations of gamma matrices $\Gamma$ and $\Gamma'$ to carry the quantum number of an axial-vector current.
Here we note that the covariant meson-field $\mathcal{H}_{v}$ and the covariant four-velocity $\mathcal{V}^{\mu}$ are used in order to make the axial-vector current be a velocity-rearrangement invariant.
We consider all the possible forms of the axial-vector currents constructed by $\Gamma=1$, $i\gamma_{5}$, $\gamma_{\mu}$, $\gamma_{\mu}\gamma_{5}$ and $\sigma_{\mu\nu}$, respectively, as follows.
The indices of isospin will be considered later.

In the case of $\Gamma=1$, there are four possible forms
\begin{eqnarray}
&& \mathrm{Tr}\, \overline{\mathcal{H}}_{v} \mathcal{H}_{v} \gamma_{\mu} \gamma_{5} 
= {\mathrm{Tr}}\, \overline{H}_{v} H_{v} \gamma_{\mu} \gamma_{5} \nonumber \\
&&\hspace{7em}
 + \frac{1}{2M} \left( {\mathrm{Tr}}\, v \!\cdot\! iD \overline{H}_{v} H_{v} \gamma_{\mu} \gamma_{5} - {\mathrm{Tr}}\, \overline{H}_{v} v \!\cdot\! iD H_{v} \gamma_{\mu} \gamma_{5} \right) \nonumber \\
&&\hspace{7em}
 + \frac{1}{4M} \varepsilon_{\mu\nu\rho\sigma} \left( {\mathrm{Tr}}\, iD^{\nu} \overline{H}_{v}H_{v} \sigma^{\rho\sigma} - {\mathrm{Tr}}\, \overline{H}_{v} iD^{\nu} H_{v} \sigma^{\rho\sigma} \right) \nonumber \\
&&\hspace{7em}
 + \mathcal{O}(1/M^2), \label{eq:axialcurrent_unit_1} \\
&& \mathrm{Tr}\, \mathcal{V}^{\mu} \overline{\mathcal{H}}_{v} \mathcal{V}^{\nu} \mathcal{H}_{v} \gamma_{\mu} \gamma_{5}
+ \mathrm{Tr}\, \mathcal{V}^{\nu} \overline{\mathcal{H}}_{v} \mathcal{V}^{\mu} \mathcal{H}_{v} \gamma_{\mu} \gamma_{5} = \mathcal{O}(1/M^2), \\
&& \varepsilon_{\mu\nu\rho\sigma} \mathrm{Tr}\, \mathcal{V}^{\rho} \overline{\mathcal{H}}_{v} \mathcal{H}_{v} \sigma^{\mu\nu}
+ \varepsilon_{\mu\nu\rho\sigma} \mathrm{Tr}\, \overline{\mathcal{H}}_{v} \mathcal{V}^{\rho} \mathcal{H}_{v} \sigma^{\mu\nu} 
= 4 \mathrm{Tr}\, \overline{\mathcal{H}}_{v} \mathcal{H}_{v} \gamma_{\sigma} \gamma_{5} + \mathcal{O}(1/M^2), \\
&& \varepsilon^{\mu\nu\rho\sigma} \mathrm{Tr} \mathcal{V}^{\lambda} \overline{\mathcal{H}}_{v} \mathcal{V}_{\nu} \mathcal{V}_{\rho} \mathcal{H}_{v} \sigma_{\mu\lambda}
+ \varepsilon^{\mu\nu\rho\sigma} \mathrm{Tr} \mathcal{V}_{\nu} \mathcal{V}_{\rho} \overline{\mathcal{H}}_{v} \mathcal{V}^{\lambda} \mathcal{H}_{v} \sigma_{\mu\lambda} 
= \mathcal{O}(1/M^2).
\end{eqnarray}
In the case of $\Gamma=i\gamma_{5}$, there are two possible forms
\begin{eqnarray}
&& \varepsilon^{\mu\nu\rho\sigma} \mathrm{Tr}\, \mathcal{V}_{\nu} \overline{\mathcal{H}}_{v} i\gamma_{5} \mathcal{V}_{\rho} \mathcal{H}_{v} \gamma_{\mu} \gamma_{5}
 + \varepsilon^{\mu\nu\rho\sigma} \mathrm{Tr}\, \mathcal{V}_{\rho} \overline{\mathcal{H}}_{v} i\gamma_{5} \mathcal{V}_{\nu} \mathcal{H}_{v} \gamma_{\mu} \gamma_{5} 
= \mathcal{O}(1/M^2), \\
&& \mathrm{Tr}\, \mathcal{V}_{\nu} \overline{\mathcal{H}}_{v} i \gamma_{5} \mathcal{H}_{v} \sigma^{\mu\nu}
+ \mathrm{Tr}\, \overline{\mathcal{H}}_{v} i \gamma_{5} \mathcal{V}_{\nu} \mathcal{H}_{v} \sigma^{\mu\nu} 
 = \mathcal{O}(1/M^2).
\end{eqnarray}
In the case of $\Gamma=\gamma_{\mu}$, there are three possible forms
\begin{eqnarray}
&& \varepsilon_{\mu\nu\rho\sigma} \mathrm{Tr}\, \overline{\mathcal{H}}_{v} \gamma^{\mu} \mathcal{H}_{v} \sigma^{\nu\rho} 
 = 2 \mathrm{Tr}\, \overline{\mathcal{H}}_{v} \mathcal{H}_{v} \gamma_{\sigma} \gamma_{5} + \mathcal{O}(1/M^2), \\
&& \mathrm{Tr}\, \overline{\mathcal{H}}_{v} \mathcal{V}\hspace{-0.6em}/ \, \mathcal{H}_{v} \gamma^{\mu}
+ \mathrm{Tr}\, \overline{\mathcal{H}}_{v} \cev{\mathcal{V}}\hspace{-0.7em}/ \, \mathcal{H}_{v} \gamma^{\mu} 
 = 2 \mathrm{Tr}\, \overline{\mathcal{H}}_{v} \mathcal{H}_{v} \gamma^{\mu} \gamma_{5} + \mathcal{O}(1/M^2), \\
&& \mathrm{Tr} \overline{\mathcal{H}}_{v} \gamma^{\mu} \mathcal{V}_{\nu} \mathcal{H}_{v} \gamma^{\nu} \gamma_{5}
+ \mathrm{Tr} \overline{\mathcal{H}}_{v} \cev{\mathcal{V}}_{\nu} \gamma^{\mu}  \mathcal{H}_{v} \gamma^{\nu} \gamma_{5} 
 = \mathcal{O}(1/M^2). \label{eq:axialcurrent_v_3}
\end{eqnarray}
In those terms, the heavy-quark-spin symmetry for $H_{v}$ is conserved.
Indeed, it is easily confirmed that those currents are invariant under the spin transformation $H_{v} \rightarrow S H_{v}$ with $S \in \mathrm{SU}(2)_{\mathrm{spin}}$ for the heavy quark.
Irrespective to the differently possible forms in Eqs.~(\ref{eq:axialcurrent_unit_1})-(\ref{eq:axialcurrent_v_3}), consequently, the axial-vector current that conserves the heavy-quark-spin symmetry is uniquely determined to be $\mathrm{Tr}\, \overline{\mathcal{H}}_{v} \mathcal{H}_{v} \gamma_{\mu} \gamma_{5}$.

For $\Gamma=\gamma^{\mu}\gamma_{5}$ and $\sigma_{\mu\nu}$,
the heavy-quark-spin symmetry is broken.
Hence, the leading term in the $1/M$-expansion in $\mathrm{Tr}\, \overline{\mathcal{H}}_{v} \Gamma \mathcal{H}_{v} \Gamma'$ with $\Gamma=\gamma^{\mu}\gamma_{5}$ and $\sigma_{\mu\nu}$ should be $\mathcal{O}(1/M^1)$ already.
This is understood from that, in the heavy quark effective theory, the terms that break the heavy-quark-spin symmetry is $\mathcal{O}(1/m_{\mathrm{Q}}^1)$ (see Eq.~(\ref{eq:HQET_Lagrangian})),
and correspondingly the terms breaking the heavy-quark-spin symmetry in the heavy meson effective theory should be also $\mathcal{O}(1/M)$.
For $\Gamma=\gamma^{\mu}\gamma_{5}$, there are six possible forms
\begin{eqnarray}
&& \mathrm{Tr}\, \overline{\mathcal{H}}_{v} \gamma^{\mu} \gamma_{5} \mathcal{H}_{v} 
 = \mathrm{Tr}\, \overline{H}_{v} \gamma^{\mu} \gamma_{5} H_{v} \nonumber \\
&&\hspace{7em}
 + \frac{1}{2M} \left( \mathrm{Tr} v \!\cdot\! iD \overline{H}_{v} \gamma^{\mu} \gamma_{5} H_{v} - \mathrm{Tr} \overline{H}_{v} \gamma^{\mu} \gamma_{5} v \!\cdot\! iD H_{v} \right) \nonumber \\
&&\hspace{7em}
 + \frac{1}{4M} \varepsilon^{\nu\mu\alpha\beta} \left( \mathrm{Tr} iD_{\nu} \overline{H}_{v} \sigma_{\alpha\beta} H_{v} - \mathrm{Tr} \overline{H}_{v} \sigma_{\alpha\beta} iD_{\nu} H_{v} \right) \nonumber \\
&&\hspace{7em}
 + \mathcal{O}(1/M^2), \\
&& \mathrm{Tr}\, \mathcal{V}^{\mu} \overline{\mathcal{H}}_{v} \mathcal{V}^{\nu} \gamma_{\nu} \gamma_{5} \mathcal{H}_{v}
+ \mathrm{Tr}\, \mathcal{V}^{\nu} \overline{\mathrm{H}}_{v} \mathcal{V}^{\mu} \gamma_{\nu} \gamma_{5} \mathcal{H}_{v}
 = \mathcal{O}(1/M^2),  \\
&& \varepsilon_{\mu\nu\rho\sigma} \mathrm{Tr}\, \mathcal{V}^{\rho} \overline{\mathcal{H}}_{v} \gamma^{\mu} \gamma_{5} \mathcal{H}_{v} \gamma^{\nu} \gamma_{5}
+ \varepsilon_{\mu\nu\rho\sigma} \mathrm{Tr}\, \overline{\mathcal{H}}_{v} \gamma^{\mu} \gamma_{5} \mathcal{V}^{\rho} \mathcal{H}_{v} \gamma^{\nu} \gamma_{5} \nonumber \\
&& = \frac{1}{M} \varepsilon_{\mu\nu\rho\sigma} \left\{ \mathrm{Tr}\, (-iD^{\rho}) \overline{H}_{v} \gamma^{\mu} \gamma_{5} H_{v} \gamma^{\nu} \gamma_{5}
 + \mathrm{Tr} \overline{H}_{v} \gamma^{\mu} \gamma_{5} iD^{\rho} H_{v} \gamma^{\nu} \gamma_{5} \right\} + \mathcal{O}(1/M^2), \\
&& \varepsilon_{\mu\nu\rho\sigma} \mathrm{Tr}\, \mathcal{V}^{\nu} \overline{\mathcal{H}}_{v} \gamma^{\mu} \gamma_{5} \mathcal{V}^{\rho} \mathcal{H}_{v} \gamma_{5}
+ \varepsilon_{\mu\nu\rho\sigma} \mathrm{Tr}\, \mathcal{V}^{\rho} \overline{\mathcal{H}}_{v} \gamma^{\mu} \gamma_{5} \mathcal{V}^{\nu} \mathcal{H}_{v} \gamma_{5}
 = \mathcal{O}(1/M^2), \\
&& \mathrm{Tr}\, \mathcal{V}^{\mu} \overline{\mathcal{H}}_{v} \gamma_{\mu} \gamma_{5} \mathcal{H}_{v} \gamma^{\nu}
+ \mathrm{Tr}\, \overline{\mathcal{H}}_{v} \mathcal{V}^{\mu} \gamma_{\mu} \gamma_{5} \mathcal{H}_{v} \gamma^{\nu}
 = \mathcal{O}(1/M^2), \\
&&\mathrm{Tr}\, \mathcal{V}_{\nu} \overline{\mathcal{H}}_{v} \gamma^{\mu} \gamma_{5} \mathcal{H}_{v} + \mathrm{Tr} \overline{\mathcal{H}}_{v} \gamma^{\mu} \gamma_{5} \mathcal{V}_{\nu} \mathcal{H}_{v} \gamma^{\nu} 
= -2 \mathrm{Tr}\, \overline{\mathcal{H}}_{v} \gamma^{\mu} \gamma_{5} \mathcal{H}_{v} + \mathcal{O}(1/M^2).
\end{eqnarray}
There are several different forms of axial-vector currents which are suppressed by $1/M$ compared to the terms, when the terms are compared in r.h.s. and l.h.s. in the above equations. 
W should note, however, that the terms in r.h.s. are $\mathcal{O}(1/M^2)$ already,
and hence they are not necessary to be considered in the desired accuracy.
Lastly, for $\Gamma=\sigma^{\mu\nu}$, there are four possible forms
\begin{eqnarray}
&& \varepsilon^{\mu\nu\rho\sigma} \mathrm{Tr}\, \overline{\mathcal{H}}_{v} \sigma_{\mu\nu} \mathcal{H}_{v} \gamma_{\rho} = 2 \mathrm{Tr}\, \overline{\mathrm{H}}_{v} \gamma^{\sigma} \gamma_{5} \mathrm{H}_{v} + \mathcal{O}(1/M^2), \\
&& \varepsilon^{\mu\nu\rho\sigma} \mathrm{Tr}\, \mathcal{V}_{\rho} \overline{\mathcal{H}}_{v} \sigma_{\mu\nu} \mathcal{H}_{v} + \varepsilon^{\mu\nu\rho\sigma} \mathrm{Tr}\, \overline{\mathrm{H}}_{v} \sigma_{\mu\nu} \mathcal{V}_{\rho} \mathcal{H}_{v} 
= - 4\mathrm{Tr}\, \overline{\mathcal{H}}_{v} \gamma^{\sigma} \gamma_{5} \mathcal{H}_{v} + \mathcal{O}(1/M^2), \\
&& \varepsilon_{\mu\rho\sigma\lambda} \mathrm{Tr}\, \overline{\mathcal{H}}_{v} \sigma^{\mu\nu} \mathcal{V}^{\rho} \mathcal{V}^{\sigma} \mathcal{H}_{v}
+ \varepsilon_{\mu\rho\sigma\lambda} \mathrm{Tr}\, \mathcal{V}^{\rho} \mathcal{V}^{\sigma} \overline{\mathcal{H}}_{v} \sigma^{\mu\nu} \mathcal{V}_{\nu} \mathcal{H}_{v} 
 = \mathcal{O}(1/M^2), \\
&& \mathrm{Tr}\, \mathcal{V}^{\mu} \overline{\mathcal{H}}_{v} \sigma_{\mu\nu} \mathcal{H}_{v} \gamma_{5}
+ \mathrm{Tr}\, \overline{\mathcal{H}}_{v} \mathcal{V}^{\mu} \mathcal{H}_{v} \gamma_{5} = \mathcal{O}(1/M^2).
\end{eqnarray}
As a conclusion, we find that the axial-vector current that breaks the heavy-quark-spin symmetry at $\mathcal{O}(1/M^1)$ is uniquely determined to be $\mathrm{Tr}\, \overline{\mathcal{H}}_{v} \gamma^{\mu} \gamma_{5} \mathcal{H}_{v}$.

As byproducts in the above calculations, we find the following axial-vector currents are $\mathcal{O}(1/M^2)$;
\begin{eqnarray}
&& \mathrm{Tr}\, v \!\cdot\! iD \overline{H}_{v} \gamma^{\mu} \gamma_{5} H_{v} - \mathrm{Tr}\, \overline{H}_{v} \gamma^{\mu} \gamma_{5} v \!\cdot\! iD H_{v},  \\
&& \varepsilon^{\nu\mu\alpha\beta} \left( \mathrm{Tr}\, iD_{\nu} \overline{H}_{v} \sigma_{\alpha\beta} H_{v} - \mathrm{Tr}\, \overline{H}_{v} \sigma_{\alpha\beta} iD_{\nu} H_{v} \right), \\
&& \varepsilon_{\mu\nu\rho\sigma} \left( \mathrm{Tr}\, (-iD^{\rho}) \overline{H}_{v} \gamma^{\mu} \gamma_{5} H_{v} \gamma^{\nu} \gamma_{5} 
 + \mathrm{Tr}\, \overline{H}_{v} \gamma^{\mu} \gamma_{5} iD^{\rho} H_{v} \gamma^{\nu} \gamma_{5} \right), \\
&& v^{\nu} \left( \mathrm{Tr}\, iD^{\mu} \overline{H}_{v} \gamma_{\mu} \gamma_{5} H_{v} - \mathrm{Tr}\, \overline{H}_{v} iD^{\mu}\gamma_{\mu} \gamma_{5} H_{v} \right), \\
&& v^{\mu}v^{\nu} \varepsilon_{\mu\rho\alpha\beta} \left( \mathrm{Tr}\, iD^{\rho} \overline{H}_{v} \sigma^{\alpha\beta} H_{v} - \mathrm{Tr}\, \overline{H}_{v} \sigma^{\alpha\beta} iD^{\rho} H_{v} \right), 
\end{eqnarray}
because they are suppressed by $1/M$ in the axial-vector currents $\mathrm{Tr}\, \overline{\mathcal{H}}_{v} \Gamma \mathcal{H}_{v} \Gamma'$ with $\Gamma=\gamma_{\mu}\gamma_{5}$ and $\sigma_{\mu\nu}$.

From the analysis above, we find that the axial-vector currents for a heavy meson are reduced to the two forms
\begin{eqnarray}
 \mathcal{A}^{(1)}_{i,\mu} &=& {\mathrm{Tr}}\, \overline{{\mathcal H}}_{v} {\mathcal H}_{v} \gamma_{\mu}\gamma_{5} \tau^{i} \nonumber \\
&=& {\mathrm{Tr}}\, \overline{H}_{v} H_{v} \gamma_{\mu} \gamma_{5} \tau^{i} \nonumber \\
&&+ \frac{1}{2M} \left( {\mathrm{Tr}}\, v \!\cdot\! iD \overline{H}_{v} H_{v} \gamma_{\mu} \gamma_{5} \tau^{i} - {\mathrm{Tr}}\, \overline{H}_{v} v \!\cdot\! iD H_{v} \gamma_{\mu} \gamma_{5} \tau^{i} \right)\nonumber \\
&&+ \frac{1}{4M} \varepsilon_{\mu\nu\rho\sigma} \left( {\mathrm{Tr}}\, iD^{\nu} \overline{H}_{v}H_{v} \sigma^{\rho\sigma} \tau^{i} - {\mathrm{Tr}}\, \overline{H}_{v} iD^{\nu} H_{v} \sigma^{\rho\sigma} \tau^{i} \right) \nonumber \\
&&+ \mathcal{O}(1/M^2),
\end{eqnarray}
and
\begin{eqnarray}
\mathcal{A}^{(2)}_{i,\mu} &=& {\mathrm{Tr}}\, \overline{{\mathcal H}}_{v} \gamma_{\mu}\gamma_{5} {\mathcal H}_{v} \tau^{i} \nonumber \\
&=& {\mathrm{Tr}}\, \overline{H}_{v} \gamma_{\mu} \gamma_{5} H_{v} \tau^{i} + \mathcal{O}(1/M^1),
\end{eqnarray}
where we introduce the Pauli matrices $\tau^{i}$ ($i=1,2,3$) for isospin.
We note that $\mathcal{A}^{(1)}_{i,\mu}$ conserves the heavy-quark-spin symmetry, while $\mathcal{A}^{(2)}_{i,\mu}$ breaks it.
From the view of the spin symmetry, therefore, $\mathcal{A}^{(1)}_{i,\mu}$ can be $\mathcal{O}(1/M^0)$ or $\mathcal{O}(1/M^1)$, while $\mathcal{A}^{(2)}_{i,\mu}$ should be $\mathcal{O}(1/M^1)$.

\subsection{Effective Lagrangian with $1/M$-corrections}

From the axial-vector currents $\mathcal{A}^{(1)}_{i,\mu}$ and $\mathcal{A}^{(2)}_{i,\mu}$ of a heavy meson, 
the effective Lagrangian up to $\mathcal{O}(1/M^1)$ is then given as
\begin{eqnarray}
&&{\mathcal L}_{\mathrm{HMET}} \nonumber \\
&=& - {\mathrm{Tr}}\, \overline{H}_{v} v \!\cdot\! iD H_{v} - {\mathrm{Tr}}\, \overline{H}_{v} \frac{(iD)^2}{2M} H_{v} + \frac{\lambda}{M} \mathrm{Tr}\, \overline{H}_{v} \sigma^{\mu\nu} H_{v} \sigma_{\mu\nu} \nonumber \\
&& + \mathrm{Tr} \left[ \left( g \mathcal{A}^{(1)}_{i,\mu} + \frac{g_{1}}{M} \mathcal{A}^{(1)}_{i,\mu} + \frac{g_{2}}{M} \mathcal{A}^{(2)}_{i,\mu} \right) a_{\perp}^{i,\mu} \right] + {\mathcal O}(1/M^2) \nonumber \\
 &=& - {\mathrm{Tr}}\, \overline{H}_{v} v \!\cdot\! iD H_{v} - {\mathrm{Tr}}\, \overline{H}_{v} \frac{(iD)^2}{2M} H_{v} \nonumber + \frac{\lambda}{M} \mathrm{Tr}\, \overline{H}_{v} \sigma^{\mu\nu} H_{v} \sigma_{\mu\nu} \\
&&+ \left( g+\frac{g_1}{M} \right) {\mathrm{Tr}}\, \overline{H}_{v} H_{v} \gamma_{\mu} \gamma_{5} a_{\perp}^{\mu} \nonumber \\
&&+ \frac{g}{2M} \left( {\mathrm{Tr}}\, v \!\cdot\! iD \overline{H}_{v} H_{v} \gamma_{\mu} \gamma_{5} a_{\perp}^{\mu} - {\mathrm{Tr}}\,  \overline{H}_{v} v \!\cdot\! iD H_{v} \gamma_{\mu} \gamma_{5} a_{\perp}^{\mu} \right) \nonumber \\
&&+ \frac{g}{4M} \varepsilon_{\mu\nu\rho\sigma} \left( {\mathrm{Tr}}\, iD^{\nu} \overline{H}_{v}H_{v} \sigma^{\rho\sigma} a_{\perp}^{\mu} - {\mathrm{Tr}}\, \overline{H}_{v} iD^{\nu} H_{v} \sigma^{\rho\sigma} a_{\perp}^{\mu} \right) \nonumber \\
&& + \frac{g_2}{M} {\mathrm{Tr}}\, \overline{H}_{v} \gamma_{\mu} \gamma_{5} H_{v} a_{\perp}^{\mu} + {\mathcal O}(1/M^2),
\label{eq:Lagrangian_HMET}
\end{eqnarray}
where $\lambda$ is related to the mass splitting between the P and P$^{\ast}$ mesons as
\begin{eqnarray}
\Delta = M_{\mathrm{P}^{\ast}} - M_{\mathrm{P}} = -8\frac{\lambda}{M}.
\end{eqnarray}
In the interaction terms, $a_{\perp}^{\mu}$ is the axial-vector current of a pion field
defined by $a_{\perp}^{\mu} = a_{\perp}^{i,\mu} \tau^{i} = \frac{i}{2}(\xi^{\dag}\partial^{\mu}\xi-\xi\partial^{\mu}\xi^{\dag})$ with $\xi=e^{i \mathcal{M}/f_{\pi}}$ and
\begin{eqnarray}
\mathcal{M} =
\frac{1}{\sqrt{2}} \vec{\pi} \cdot \vec{\tau} =
\left(
\begin{array}{cc}
 \frac{\pi^{0}}{\sqrt{2}} & \pi^{+}  \\
 \pi^{-} & -\frac{\pi^{0}}{\sqrt{2}} 
\end{array}
\right),
\end{eqnarray}
(a sum is taken over $i=1,2,3$).
$f_{\pi}=135$ MeV is the pion decay constant.
Here $g$, $g_1$ and $g_2$ are unknown coupling constants.
The effective Lagrangian in Eq.~(\ref{eq:Lagrangian_HMET}) coincides with the one given in Ref.~\cite{Kitazawa:1993bk}.
We can easily confirm that the effective Lagrangian is invariant up to $\mathcal{O}(1/M^1)$ under the velocity-rearrangement in Eq.~(\ref{eq:four_velocity_change}).

The coupling constants $g$, $g_1$ and $g_2$ should be fixed from theoretical calculations or from experimental information.
At $\mathcal{O}(1/M^{0})$ in the heavy quark limit, the value of $g$ has been discussed by several theoretical approaches such as the quark models \cite{Yan:1992gz,Colangelo:1994jc,Becirevic:1999fr}, the QCD sum rules \cite{Colangelo:1994es,Belyaev:1994zk,Dosch:1995kw,Colangelo:1997rp,Wang:2006ida}, the lattice QCD simulations \cite{deDivitiis:1998kj,Abada:2003un,Negishi:2006sc,Ohki:2008py,Becirevic:2009yb,Bulava:2010ej,Bulava:2011yz,Detmold:2011bp,Detmold:2012ge}, analyses of weak decays of B mesons \cite{Cho:1992nt,Arnesen:2005ez,Li:2010rh} and analyses of strong decays of $\Sigma_{\mathrm{c}}^{\ast}$ baryons \cite{Cheng:1997rp}.
In the present study, we use $g=0.4-0.5$ which are consistent with the results in the lattice QCD simulations as summarized in Ref.~\cite{Detmold:2012ge}.

To determine $g_{1}$ and $g_{2}$, we impose a constraint to reproduce the observed decay width of $\mathrm{D}^{\ast} \rightarrow \mathrm{D}\pi$.
The decay width is given at tree level from Eq.~(\ref{eq:Lagrangian_HMET}) as 
\begin{eqnarray}
\hspace{-1.5em} \Gamma = \frac{4\pi}{32\pi^2} \frac{1}{f_{\pi}^{2}} \frac{1}{3} \left[ 2\left( g + \frac{g_{1}}{M}-\frac{g_{2}}{M} \right) + \frac{g}{M} v\cdot p_{\pi} \right]^2 |\vec{p}_{\pi}|^3,
\label{eq:decay_width}
\end{eqnarray}
where $v^{\mu}=(1,\vec{0}\,)$ is the four-velocity of the $\mathrm{D}^{\ast}$ meson in the initial state at rest, and $p_{\pi}^{\,\mu}=(\sqrt{\vec{p}_{\pi}^{\,\,2} + m_{\pi}^{2}},\vec{p}_{\pi})$ is the four-momentum of the pion in the final state.
When Eq.~(\ref{eq:decay_width}) is compared with the case in the heavy quark limit, the terms of $g/M$, $g_{1}/M$ and $g_{2}/M$ are new ingredients as the $1/M$-corrections.
From the experimental value $\Gamma=0.065$ MeV for $\mathrm{D}^{\ast +} \rightarrow \mathrm{D}^{0} \pi^{+}$ \cite{Beringer:1900zz}, we obtain possible combinations of $g_{1}$ and $g_{2}$ for a given $g$.
In most of our presentation, we set $g_1=0$. 
It will turn out that our conclusion does not affected qualitatively so much even for the case of $g_1 \neq 0$, as discussed later.
We consider the two parameter sets; $(g,g_{1}/M_{\mathrm{D}},g_{2}/M_{\mathrm{D}})=(0.5,0,-0.07)$ and $(0.4,0,-0.18)$.
Note that each $g_1$ and $g_2$ has a dimension of energy.
It will be useful to introduce a dimensionless number $g_{1}/M_{\mathrm{D}}$ and $g_{2}/M_{\mathrm{D}}$ as $g_{1}$ and $g_{2}$ are divided by some quantity with a dimension of energy, say the mass $M_{\mathrm{D}}$ of a D meson.

\section{Self-energies of $\bar{\mathrm{D}}^{(\ast)}$ and $\mathrm{B}^{(\ast)}$ mesons in nuclear matter}
\label{sec:nuclear_matter}

Based on the heavy meson effective Lagrangian (\ref{eq:Lagrangian_HMET}), we consider the self-energies of $\bar{\mathrm{D}}^{(\ast)}$ and $\mathrm{B}^{(\ast)}$ mesons in nuclear matter.
We discuss first the $\bar{\mathrm{D}}^{(\ast)}$ meson, then apply the similar discussion to the $\mathrm{B}^{(\ast)}$ meson only by changing  the mass $M_{\bar{\mathrm{D}}^{(\ast)}}$ to $M_{\mathrm{B}^{(\ast)}}$ with keeping the values of $g$, $g_{1}$ and $g_{2}$.
We have so far used the meson-field $P^{(\ast)}$ for the $\mathrm{Q}\bar{\mathrm{q}}$ meson in the Lagrangian (\ref{eq:Lagrangian_HMET}).
We easily obtain the interaction vertex with pion in the $\bar{\mathrm{Q}}\mathrm{q}$ meson sector by changing the sign of the interaction vertex in Eq.~(\ref{eq:Lagrangian_HMET}).

\begin{figure}[tbp]
\includegraphics[width=6cm,angle=0]{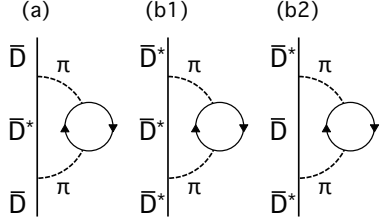}
\caption{The diagrams of the self-energies for (a) the $\bar{\mathrm{D}}$ meson and (b1,b2) the $\bar{\mathrm{D}}^{\ast}$ meson in the nuclear matter. The thick solid (normal dashed) lines denote the propagator of the $\bar{\mathrm{D}}$ and $\bar{\mathrm{D}}^{\ast}$ mesons (pions). The thin solid lines with arrows indicate the nucleon and hole propagators.}
\label{fig:selfenergy}
\end{figure}

We consider the self-energies of the $\bar{\mathrm{D}}^{(\ast)}$ mesons with the lowest order of pion loops whose diagrams are shown in Fig.~\ref{fig:selfenergy}. 
Those diagrams were considered in our previous works \cite{Yasui:2012rw}, where the complete $1/M$-expansion was not given yet due to the lack of the $1/M$-corrections in the interaction vertices.
The self-energy of the $\bar{\mathrm{D}}$ meson in Fig.~\ref{fig:selfenergy}(a) is given by
\begin{eqnarray}
&& -i \Sigma_{\bar{\mathrm{D}}}(\rho;v,k) \nonumber \\
&=& -\frac{2}{f_{\pi^2}} \int \frac{\mathrm{d}^4 \ell}{(2\pi)^4}
\frac{\left\{ g^2 + 2g \left( g_1 - g_2 \right)/M \right\} \ell^2 - g^2 (v \!\cdot\! \ell/M) \left\{ (2k-\ell) \!\cdot\! \ell - v \!\cdot\! (2k-\ell) v\!\cdot\! \ell \right\}}{2 v \!\cdot\! (k-\ell) + (k-\ell)^2/M - 2\Delta + i\eta} \nonumber \\
&& \times \frac{1}{(\ell^2-m_{\pi}^2+i\eta)^2} \tau^{a} \Pi_{\pi}^{ab}(\ell) \tau^{b},
\label{eq:selfenergy_D}
\end{eqnarray}
 with the internal momentum $\ell^{\mu}$ carried by pions.
Here we measure the self-energy from the mass of $\bar{\mathrm{D}}$ meson in vacuum.
For this, we have transformed $H_{v}(x) \rightarrow e^{i(3/4)\Delta v \cdot x} H_{v}(x)$, which gives only a mass shift in Eq.~(\ref{eq:Lagrangian_HMET}).
We define the self-energy $-i\Pi_{\pi}^{ab}(\rho;\ell)$ of the pion propagating with momentum $\ell$, which is given by the lowest order of pion loops as
\begin{eqnarray}
-i\Pi_{\pi}^{ab}(\rho;\ell) = -i\Pi_{\pi}^{(1)ab}(\rho;\ell)-i\Pi_{\pi}^{(2)ab}(\rho;\ell),
\end{eqnarray}
with
\begin{eqnarray}
-i\Pi_{\pi}^{(1)ab}(\rho;\ell) =&& 
\left( -\frac{g_{\mathrm{A}}}{2f_{\pi}}i \right)^2 
(-1) \mathrm{Tr} \int \frac{\mathrm{d}^{4}p_{1}}{(2\pi)^4} i \ell \hspace{-0.5em}/ \gamma_{5} \tau^{a}
 \frac{i}{p_{1}\hspace{-0.8em}/+\ell\hspace{-0.5em}/-m_{\mathrm{N}}+i\eta}
 (-i\ell\hspace{-0.5em}/ \gamma_{5}) \tau^{b} \nonumber \\
&& \times (p_{1}\hspace{-0.8em}/ + m_{\mathrm{N}}) (-2\pi) \delta(p_{1}^{2}-m_{\mathrm{N}}^{2}) \theta(p_{10}) \theta(k_{\mathrm{F}}-|\vec{p}_{1}|),
\end{eqnarray}
and
\begin{eqnarray}
-i\Pi_{\pi}^{(2)ab}(\rho;\ell) = &&
\left( -\frac{g_{\mathrm{A}}}{2f_{\pi}}i \right)^2
(-1) \mathrm{Tr} \int \frac{\mathrm{d}^{4}p_{1}}{(2\pi)^4} i \ell \hspace{-0.5em}/ \gamma_{5} \tau^{a}
(p_{1}\hspace{-0.8em}/+\ell\hspace{-0.5em}/+m_{\mathrm{N}}) (-2\pi) \delta\left( (p_{1}+\ell)^2-m_{\mathrm{N}}^2 \right) \nonumber \\
&& \times \theta(p_{10}+\ell_{0}) \theta(k_{\mathrm{F}}-|\vec{p}_{1}+\vec{\ell}\,|)
 (-i\ell\hspace{-0.5em}/ \gamma_{5}) \tau^{b}
 (p_{1}\hspace{-0.8em}/ + m_{\mathrm{N}}) \nonumber \\
&& (-2\pi) \delta(p_{1}^{2}-m_{\mathrm{N}}^{2}) \theta(p_{10}) \theta(k_{\mathrm{F}}-|\vec{p}_{1}|).
\end{eqnarray}
Here we have used the axial-vector current coupling for the NN$\pi$ vertex
\begin{eqnarray}
{\cal L}_{\mathrm{N}\mathrm{N}\pi} = \frac{g_{\mathrm{A}}}{2f_{\pi}} \bar{N} \gamma_{\mu} \gamma_{5} \partial^{\mu} \vec{\pi} \!\cdot\! \vec{\tau} N,
\end{eqnarray}
with the coupling constant $g_{\mathrm{A}}=1.3$,
and the in-medium propagator
\begin{eqnarray}
 (p\hspace{-0.4em}/+m_{\mathrm{N}})
\left[ \frac{i}{p^{2}-m_{\mathrm{N}}^2+i\eta} 
 -2\pi \delta(p^2-m_{\mathrm{N}}^2) \theta(p_{0}) \theta(k_{\mathrm{F}}-|\vec{p}\,|) \right]  \mathbf{1}_{\mathrm{f}},
\label{eq:nucleon_propagator}
\end{eqnarray}
 for the nucleon (mass $m_{\mathrm{N}}$) carrying the four-momentum $p^{\mu}=(p_{0},\vec{p}\,)$, and the $2 \times 2$ unitary matrix $\mathbf{1}_{\mathrm{f}}$ for isospin space in isospin-symmetric nuclear matter with Fermi momentum $k_{\mathrm{F}}$.
Here $\eta$ is an infinitely small positive number.
The second term in the square brackets in Eq.~(\ref{eq:nucleon_propagator}) indicates to subtract the on-mass-shell nucleon states with positive energies inside the Fermi surface, because these states are not allowed to propagate due to the Pauli blocking effects.
The baryon number density is given by $\rho=2k_{\mathrm{F}}^3/3\pi^2$.
We comment that the vector current coupling with two pions is not considered in the present discussion.
This can be justified, because their contribution to the self-energy of P$^{(\ast)}$ meson vanishes in isospin-symmetric nuclear matter.

The self-energy of the $\bar{\mathrm{D}}^{\ast}$ meson in nuclear matter is given by
\begin{eqnarray}
-i \Sigma_{\bar{\mathrm{D}}^{\ast}}(\rho;v,k) = 
-i \Sigma_{\bar{\mathrm{D}}^{\ast}}^{(\bar{\mathrm{D}})}(\rho;v,k) -i \Sigma_{\bar{\mathrm{D}}^{\ast}}^{(\bar{\mathrm{D}}^{\ast})}(\rho;v,k),
\label{eq:selfenergy_D*}
\end{eqnarray}
with
\begin{eqnarray}
&& -i \Sigma_{\bar{\mathrm{D}}^{\ast}}^{(\bar{\mathrm{D}})}(\rho;v,k) \nonumber \\
&=& -\frac{2}{f_{\pi^2}} \int \frac{\mathrm{d}^4 \ell}{(2\pi)^4}
\frac{\left\{ g^2 + 2g \left( g_1 + g_2 \right)/M \right\} \ell^2 - g^2 (v \!\cdot\! \ell/M) \left\{ (2k-\ell) \!\cdot\! \ell - v \!\cdot\! (2k-\ell) v\!\cdot\! \ell \right\}}{2 v \!\cdot\! (k-\ell) + (k-\ell)^2/M + i\eta}
\nonumber \\
&& \times \frac{1}{(\ell^2-m_{\pi}^2+i\eta)^2} \tau^{a} \Pi_{\pi}^{ab}(\ell) \tau^{b},
\label{eq:integral_b1}
\end{eqnarray}
and
\begin{eqnarray}
&& -i \Sigma_{\bar{\mathrm{D}}^{\ast}}^{(\bar{\mathrm{D}}^{\ast})}(\rho;v,k) \nonumber \\
&=& -\frac{2}{f_{\pi^2}} \int \frac{\mathrm{d}^4 \ell}{(2\pi)^4}
\frac{\left\{ g^2 + 2g \left( g_1 - g_2 \right)/M \right\} \ell^2 - g^2 (v \!\cdot\! \ell/M) \left\{ (2k-\ell) \!\cdot\! \ell - v \!\cdot\! (2k-\ell) v\!\cdot\! \ell \right\}}{2 v \!\cdot\! (k-\ell) + (k-\ell)^2/M + 2\Delta + i\eta} \nonumber \\
&& \times \frac{1}{(\ell^2-m_{\pi}^2+i\eta)^2} \tau^{a} \Pi_{\pi}^{ab}(\ell) \tau^{b},
\label{eq:integral_b2}
\end{eqnarray}
from (b1) and (b2) in Fig.~\ref{fig:selfenergy}, respectively.
Here we measure the self-energy from the mass of $\bar{\mathrm{D}}^{\ast}$ meson in vacuum.
For this, we have transformed $H_{v}(x) \rightarrow e^{-i(1/4)\Delta v \cdot x} H_{v}(x)$, which gives only a mass shift in Eq.~(\ref{eq:Lagrangian_HMET}).
As a matter of fact, each of the self-energies of the $\bar{\mathrm{D}}$ and $\bar{\mathrm{D}}^{\ast}$ mesons in Eqs.~(\ref{eq:selfenergy_D}) and (\ref{eq:selfenergy_D*}) is invariant under the velocity-rearrangement in Eq.~(\ref{eq:four_velocity_change}).

In the integrals in Eqs.~(\ref{eq:selfenergy_D}) and (\ref{eq:selfenergy_D*}), we first perform the $\ell_{0}$-integrals to give the three dimensional integrals.
Because the three-dimensional integrals are divergent, we introduce the cutoff regularization by introducing a momentum cutoff parameter $\Lambda$, as it was done in our previous work in Ref.~\cite{Yasui:2012rw} (see also Refs.~\cite{Kaiser:1997mw,Kaiser:2001jx,Kaiser:2001ra,Kaiser:2004fe,Kaiser:2007au,Finelli:2009tu} for discussions in nuclear and hypernuclear matter):
\begin{eqnarray}
&& \mathcal{P} \int \frac{\mathrm{d}^{3} \vec{\ell}}{(2\pi)^{3}}
\frac{|\vec{\ell}|^{4}}{x+\frac{1}{2}(1+\frac{m_{\mathrm{N}}}{M}) |\vec{\ell}|^{\,2} + \vec{p}_{1} \!\cdot\! \vec{\ell} - i\eta} \hspace{0.2em} \frac{1}{(|\vec{\ell}|^{\,2} + m_{\pi}^{2})^{2}} \nonumber \\
&\rightarrow &
\mathcal{P} \int \frac{\mathrm{d}^{3} \vec{\ell}}{(2\pi)^{3}}
\frac{|\vec{\ell}|^{4}}{x+\frac{1}{2}(1+\frac{m_{\mathrm{N}}}{M}) |\vec{\ell}|^{\,2} + \vec{p}_{1} \!\cdot\! \vec{\ell} - i\eta} \hspace{0.2em} \frac{1}{(|\vec{\ell}|^{\,2} + m_{\pi}^{2})^{2}}
- \int \frac{\mathrm{d}^{3} \vec{\ell}}{(2\pi)^{3}}
\frac{1}{\frac{1}{2}(1+\frac{m_{\mathrm{N}}}{M}) |\vec{\ell}|^{\,2}} \nonumber \\
&&
+\int_{|\vec{\ell}| \le \Lambda} \frac{\mathrm{d}^{3} \vec{\ell}}{(2\pi)^{3}}
\frac{1}{\frac{1}{2}(1+\frac{m_{\mathrm{N}}}{M}) |\vec{\ell}|^{\,2}},
\end{eqnarray}
with $x=0$, $\pm m_{\mathrm{N}}\Delta$.
Here, $\mathcal{P}$ stands for the principal value integration.
The sum of the first and second integrals becomes finite in the integration at $|\vec{\ell}| \rightarrow \infty$, while the last term becomes finite by introducing the cutoff parameter $\Lambda$.
Thus, we succeed to separate the cutoff-independent (first and second) term and the cutoff-dependent (third) term.
The cutoff parameters are set to be
$\Lambda_{\bar{\mathrm{D}}} = 1.27 \Lambda$ and
$\Lambda_{\mathrm{B}} = 1.22 \Lambda$ with $\Lambda=700$ MeV
for $\bar{\mathrm{D}}^{(\ast)}$ and $\mathrm{B}^{(\ast)}$ mesons, which were fixed from the analysis of the hadron sizes as discussed in Ref.~\cite{Yasui:2012rw}.

From the self-energies in Eqs.~(\ref{eq:selfenergy_D}) and (\ref{eq:selfenergy_D*}), the in-medium masses of the $\bar{\mathrm{D}}^{(\ast)}$ and $\mathrm{B}^{(\ast)}$ mesons in nuclear matter are given as
\begin{eqnarray}
M_{\bar{\mathrm{D}}}(\rho) &=& M_{\bar{\mathrm{D}}} + \Sigma_{\bar{\mathrm{D}}}(\rho;v_{\mathrm{r}},0), \label{eq:inmedium_mass_D} \\
M_{\bar{\mathrm{D}}^{\ast}}(\rho) &=& M_{\bar{\mathrm{D}}^{\ast}} + \Sigma_{\bar{\mathrm{D}}^{\ast}}(\rho;v_{\mathrm{r}},0), \label{eq:inmedium_mass_D*} \\
M_{\mathrm{B}}(\rho) &=& M_{\mathrm{B}} + \Sigma_{\mathrm{B}}(\rho;v_{\mathrm{r}},0), \label{eq:inmedium_mass_B} \\
M_{\mathrm{B}^{\ast}}(\rho) &=& M_{\mathrm{B}^{\ast}} + \Sigma_{\mathrm{B}^{\ast}}(\rho;v_{\mathrm{r}},0). \label{eq:inmedium_mass_B*}
\end{eqnarray}
in the rest frame with $v_{\mathrm{r}}=(1,\vec{0}\,)$ and $k^{\mu}=0$.
$M_{\bar{\mathrm{D}}^{(\ast)}}$ and $M_{\mathrm{B}^{(\ast)}}$ are masses of $\bar{\mathrm{D}}^{(\ast)}$ and $\mathrm{B}^{(\ast)}$ mesons in vacuum.
We consider the normal nuclear matter with the baryon number density $\rho_{0}=0.17$ fm$^{-3}$ ($k_{\mathrm{F}}=270$ MeV).
As results, we obtain the in-medium masses; $M_{\bar{\mathrm{D}}}(\rho_{0})=1851.2$ MeV, $M_{\bar{\mathrm{D}}^{\ast}}(\rho_{0})=1995.9-i65.5$ MeV, $M_{\mathrm{B}}(\rho_{0})=5237.3$ MeV and $M_{\mathrm{B}^{\ast}}(\rho_{0})=5254.1-i43.0$ MeV for $(g,g_{1}/M_{\mathrm{D}},g_{2}/M_{\mathrm{D}})=(0.5, 0, -0.07)$,
 and $M_{\bar{\mathrm{D}}}(\rho_{0})=1852.3$ MeV, $M_{\bar{\mathrm{D}}^{\ast}}(\rho_{0})=1978.9-i47.6$ MeV, $M_{\mathrm{B}}(\rho_{0})=5247.3$ MeV and $M_{\mathrm{B}^{\ast}}(\rho_{0})=5281.1-i28.0$ MeV for $(g,g_{1}/M_{\mathrm{D}},g_{2}/M_{\mathrm{D}})=(0.4, 0, -0.17)$.
Concerning the $\bar{\mathrm{D}}^{\ast}$ ($\mathrm{B}^{\ast}$) meson, the imaginary part
 gives the width in the decay process from the $\bar{\mathrm{D}}^{\ast}$ ($\mathrm{B}^{\ast}$) meson to the $\bar{\mathrm{D}}$ ($\mathrm{B}$) meson in nuclear matter.
However, they are irrelevant to the present discussion, because only the real parts are important as the mass shifts.
In any cases, the real parts of the in-medium masses of $\bar{\mathrm{D}}^{(\ast)}$ and $\mathrm{B}^{(\ast)}$ mesons are smaller by a few ten MeV than the masses in vacuum.
This means that $\bar{\mathrm{D}}^{(\ast)}$ and $\mathrm{B}^{(\ast)}$ mesons are bound in the normal nuclear matter.

By fitting those masses to the in-medium mass formula in Eq.~(\ref{eq:mass_formula_density}),
we obtain the matrix elements $\bar{\Lambda}(\rho_{0})$, $\lambda_{1}(\rho_{0})$ and $\lambda_{2}(\rho_{0};m_{\mathrm{Q}})$ ($m_{\mathrm{Q}}=m_{\mathrm{c}}$ and $m_{\mathrm{b}}$) in the normal nuclear matter.
In Eq.~(\ref{eq:mass_formula_density}), there is an operator $\vec{S}_{\mathrm{L}}$ for the total angular momentum of the light component.
In the present case, the light component for a $\bar{\mathrm{D}}^{(\ast)}$ ($\mathrm{B}^{(\ast)}$) meson includes, not only light quarks and gluons inside the meson, but also the nucleon-hole pairs in the nuclear matter.
Such a complex structure of light quarks, gluons and nucleon-holes is called the spin-complex \cite{Yasui:2013vca}.
As for the quantum number, the spin-complex (light component) for a $\bar{\mathrm{D}}^{(\ast)}$ ($\mathrm{B}^{(\ast)}$) meson in nuclear matter should have isospin, total angular momentum and parity $I(j^{\mathcal{P}})=1/2(1/2^{+})$, as discussed in Ref.~\cite{Yasui:2013vca}.
Therefore, we have $S_{\mathrm{L}}=1/2$ for the $\bar{\mathrm{D}}^{(\ast)}$ ($\mathrm{B}^{(\ast)}$) meson in nuclear matter.
Finally by using $m_{\mathrm{c}}=1.30$ GeV and $m_{\mathrm{b}}=4.71$ GeV \cite{Neubert:1993mb},
we obtain the results summarized in Table~\ref{table:D_B_medium}.
The matrix elements are $\bar{\Lambda}(\rho_0)=0.51$ GeV, $\lambda_{1}(\rho_0)=-0.32$ GeV$^{2}$, $\lambda_{2}(\rho_0;m_{\mathrm{c}})=0.068$ GeV$^{2}$ and $\lambda_{2}(\rho_0;m_{\mathrm{b}})=0.039$ GeV$^{2}$ for $(g,g_{1}/M_{\mathrm{D}},g_{2}/M_{\mathrm{D}})=(0.5,0,-0.07)$, and
$\bar{\Lambda}(\rho_0)=0.53$ GeV, $\lambda_{1}(\rho_0)=-0.30$ GeV$^{2}$, $\lambda_{2}(\rho_0;m_{\mathrm{c}})=0.082$ GeV$^{2}$ and $\lambda_{2}(\rho_0;m_{\mathrm{b}})=0.080$ GeV$^{2}$ for $(g,g_{1}/M_{\mathrm{D}},g_{2}/M_{\mathrm{D}})=(0.4,0,-0.17)$.

Concerning $\bar{\mathrm{D}}^{(\ast)}$ ($\mathrm{B}^{(\ast)}$) mesons in vacuum for comparison, 
we obtain the matrix elements $\bar{\Lambda}=0.58$ GeV, $\lambda_{1}=-0.25$ GeV$^{2}$, $\lambda_{2}(m_{\mathrm{c}})=0.092$ GeV$^{2}$ and $\lambda_{2}(m_{\mathrm{b}})=0.11$ GeV$^{2}$, as summarized in the last row in Table.~\ref{table:D_B_medium}.
Those values are very close to the ones obtained by the analysis of the QCD sum rules: $\bar{\Lambda}=0.57 \pm 0.07$ GeV, $\lambda_{1}=-0.25 \pm 0.20$ GeV$^{2}$ and $\lambda_{2}=0.12 \pm 0.02$ GeV$^{2}$ \cite{Neubert:1993mb}.
It is confirmed that, when the matrix elements calculated in nuclear matter at $\rho=0$ in Eqs.~(\ref{eq:inmedium_mass_D})-(\ref{eq:inmedium_mass_B*}), the values of the matrix elements coincide with those in vacuum; $\bar{\Lambda}(0)=\bar{\Lambda}$, $\lambda_{1}(0)=\lambda_{1}$ and $\lambda_{2}(0;m_{\mathrm{Q}})=\lambda_{2}(m_{\mathrm{Q}})$.

Let us compare the matrix elements in nuclear matter with the ones in vacuum.
We find that $\bar{\Lambda}(\rho_0)$ is reduced by $0.87-0.91$ times than $\bar{\Lambda}$ in vacuum.
We also see that $\lambda_{1}(\rho_{0})$ is enhanced by $1.28-1.20$ times in the absolute values than $\lambda_{1}$ in vacuum and $\lambda_{2}(\rho_0;m_{\mathrm{c}})$ and $\lambda_{2}(\rho_{0};m_{\mathrm{b}})$ are reduced by $0.74-0.89$ times and $0.35 - 0.73$ times, respectively, than $\lambda_{2}(m_{\mathrm{c}})$ and $\lambda_{2}(m_{\mathrm{b}})$ in vacuum.

\begin{figure}[tbp]
\includegraphics[width=8cm,angle=0]{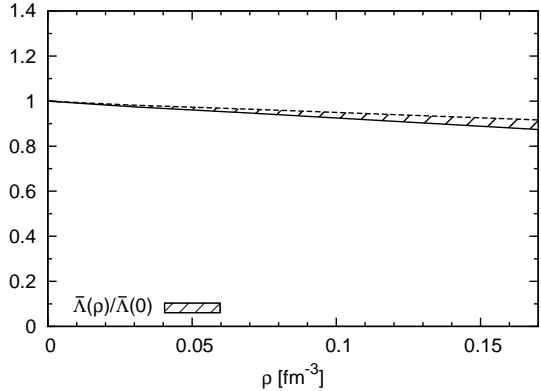}
\caption{The ratio $\bar{\Lambda}(\rho)/\bar{\Lambda}(0)$ for $\bar{\mathrm{D}}^{(\ast)}$ and $\mathrm{B}^{(\ast)}$ mesons in nuclear matter as a function of the baryon number density $\rho$. The solid and dashed lines indicate the results for $(g, g_{1}/M_{\mathrm{D}}, g_{2}/M_{\mathrm{D}})=(0.5, 0, -0.07)$ and $(0.4, 0, -0.17)$, respectively.}
\label{fig:lambda_bar}
\end{figure}

\begin{figure}[tbp]
\includegraphics[width=8cm,angle=0]{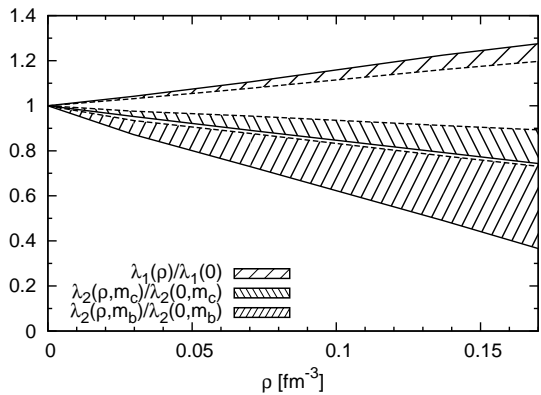}
\caption{The ratios $\lambda_{1}(\rho)/\lambda_{1}(0)$, $\lambda_{2}(\rho;m_{\mathrm{c}})/\lambda_{2}(0;m_{\mathrm{c}})$ and $\lambda_{2}(\rho;m_{\mathrm{b}})/\lambda_{2}(0;m_{\mathrm{b}})$ (from top to down) with the same notations in Fig.~\ref{fig:lambda_bar}.}
\label{fig:lambda}
\end{figure}

By utilizing 
the matrix elements $\bar{\Lambda}(\rho)$, $\lambda_{1}(\rho)$, $\lambda_{2}(\rho;m_{\mathrm{c}})$ and $\lambda_{2}(\rho;m_{\mathrm{b}})$,
 we get the information about the change of the gluon fields in the nuclear matter from those in vacuum.
From Eqs.~(\ref{eq:ratio_lambda}), (\ref{eq:ratio1}) and (\ref{eq:ratio2}), we obtain the modifications of gluon fields in the nuclear matter,
\begin{eqnarray}
\frac{\bar{\Lambda}(\rho)}{\bar{\Lambda}} &=& \frac{\langle \tilde{\mathrm{H}}_{v_{\mathrm{r}}}(\rho) | \frac{\beta(\alpha_{\mathrm{s}})}{4\alpha_{\mathrm{s}}} G^2 | \tilde{\mathrm{H}}_{v_{\mathrm{r}}}(\rho) \rangle}
 {\langle \tilde{\mathrm{H}}_{v_{\mathrm{r}}} | \frac{\beta(\alpha_{\mathrm{s}})}{4\alpha_{\mathrm{s}}} G^2 | \tilde{\mathrm{H}}_{v_{\mathrm{r}}} \rangle} \nonumber \\
 & =& 0.87-0.91, \\
\frac{\lambda_{1}(\rho)}{\lambda_{1}} &=& \frac{\langle \mathrm{H}_{v_{\mathrm{r}}}(\rho) | \overline{Q}_{v_{\mathrm{r}}} g_{\mathrm{s}} \vec{x} \!\cdot\! \vec{E} Q_{v_{\mathrm{r}}} | \mathrm{H}_{v_{\mathrm{r}}}(\rho) \rangle}{\langle \mathrm{H}_{v_{\mathrm{r}}} | \overline{Q}_{v_{\mathrm{r}}} g_{\mathrm{s}} \vec{x} \!\cdot\! \vec{E} Q_{v_{\mathrm{r}}} | \mathrm{H}_{v_{\mathrm{r}}} \rangle} \nonumber \\
 &=& 1.28-1.20 \\
\frac{\lambda_{2}(\rho;m_{\mathrm{Q}})}{\lambda_{2}(m_{\mathrm{Q}})} &=&  \frac{\langle \mathrm{H}_{v_{\mathrm{r}}}(\rho) | \overline{Q}_{v_{\mathrm{r}}} g_{\mathrm{s}} \vec{\sigma} \!\cdot\! \vec{B} Q_{v_{\mathrm{r}}} | \mathrm{H}_{v_{\mathrm{r}}}(\rho) \rangle}{\langle \mathrm{H}_{v_{\mathrm{r}}} | \overline{Q}_{v_{\mathrm{r}}} g_{\mathrm{s}} \vec{\sigma} \!\cdot\! \vec{B} Q_{v_{\mathrm{r}}} | \mathrm{H}_{v_{\mathrm{r}}} \rangle} \nonumber \\
 &=&
\left\{
\begin{array}{l}
0.74-0.89 \hspace{1em} \mathrm{(charm)} \\
0.35-0.73 \hspace{1em} \mathrm{(bottom)}
\end{array}
\right. ,
\end{eqnarray}
at $\rho=\rho_{0}$.
The first equation indicates that the effect of scale anomaly is suppressed in the nuclear matter.
The second and third equations indicate that the contribution to the in-medium masses from the chromoelectric gluon is enhanced, while that from the chromomagnetic gluon is reduced.
The tendency that $\lambda_1(\rho_0)$ is enhanced and $\bar{\Lambda}(\rho_0)$, $\lambda_2(\rho_0;m_{\mathrm{c}})$ and $\lambda_2(\rho_0;m_{\mathrm{b}})$ are reduced is not affected by the small change of the parameter $(g,g_{1}/M_{\mathrm{D}},g_{2}/M_{\mathrm{D}})$.
Such tendencies are seen also for another baryon number densities different from $\rho_{0}$.
We plot the ratio $\bar{\Lambda}(\rho)/\bar{\Lambda}$ as a function of $\rho$ in Fig.~\ref{fig:lambda_bar}, and
 the ratios $\lambda_{1}(\rho)/\lambda_{1}$, $\lambda_{2}(\rho;m_{\mathrm{c}})/\lambda_{2}(m_{\mathrm{c}})$ and $\lambda_{2}(\rho;m_{\mathrm{b}})/\lambda_{2}(m_{\mathrm{b}})$ in Fig.~\ref{fig:lambda}.
With the parameter sets $(g,g_{1}/M_{\mathrm{D}},g_{2}/M_{\mathrm{D}})=(0.5,0,-0.07)$ and $(0.4,0,-0.18)$, we confirm that $\lambda_{1}(\rho)$ becomes enhanced, while $\bar{\Lambda}(\rho)$, $\lambda_{2}(\rho;m_{\mathrm{c}})$ and $\lambda_{2}(\rho;m_{\mathrm{b}})$ become reduced, as the baryon number density increases.

It may be interesting to compare $\bar{\Lambda}(\rho)$
 with the gluon condensate in nuclear matter, because both of them are related to the scale anomaly in QCD.
In Ref.~\cite{Cohen:1991nk,Cohen:1994wm}, it was discussed that the gluon condensate in nuclear matter at normal density ($\rho=\rho_{0}$) is, roughly to say, about 0.95 of that in the QCD vacuum.
This is comparable with the number $0.87-0.91$ for $\bar{\Lambda}(\rho_{0})/\bar{\Lambda}$ in our results.
It would be natural to have similar numbers for both of them, because their origins are the same; the scale anomaly in QCD.
We should keep in mind, however, that this comparison is done only qualitatively, since we do not take into account the terms linear to light quark current masses, namely the contributions from scalar quark condensates, in Eqs.~(\ref{eq:lambda}) and (\ref{eq:lambda_density}).
Concerning $\lambda_{1}$ and $\lambda_{2}(m_{\mathrm{Q}})$, their enhancement and reduction in nuclear matter are new phenomena which seem to have not been addressed in the literature.
The enhancement of $\lambda_{1}$ in nuclear matter would be reasonable.
Because the the kinetic energy of the meson (see Eq.~(\ref{eq:chromoelectric0})) should increase, when the meson is bound in nuclear matter.
The reduction of $\lambda_{2}(m_{\mathrm{Q}})$ in nuclear matter is interesting.
It suggests that the mass splitting between $\bar{\mathrm{D}}$ and $\bar{\mathrm{D}}^{\ast}$ mesons ($\mathrm{B}$ and $\mathrm{B}^{\ast}$ mesons) becomes small in nuclear matter.

\begin{table}[tbp]
\caption{The matrix elements $\bar{\Lambda}(\rho)$, $\lambda_{1}(\rho)$, $\lambda_{2}(\rho,m_{\mathrm{c}})$ and $\lambda_{2}(\rho,m_{\mathrm{b}})$ of $\bar{\mathrm{D}}^{(\ast)}$ and $\mathrm{B}^{(\ast)}$ mesons in normal nuclear matter ($\rho=\rho_{0}$) with $g_{1}=0$. The last row indicates the values in vacuum ($\rho=0$).}
\begin{center}
\begin{tabular}{|c|c|c|c|c|}
\hline
  $(g, g_{1}/M_{\mathrm{D}}, g_{2}/M_{\mathrm{D}})$ &  $\bar{\Lambda}(\rho)$ [GeV] & $\lambda_{1}(\rho)$ [GeV$^2$] & $\lambda_{2}(\rho;m_{\mathrm{c}})$ [GeV$^2$] & $\lambda_{2}(\rho;m_{\mathrm{b}})$ [GeV$^2$] \\
\hline
 $(0.5, 0, -0.07)$ & 0.51 & -0.32 & 0.068 & 0.039 \\
 $(0.4, 0, -0.17)$ & 0.53 & -0.30 & 0.082 & 0.080 \\
\hline
vacuum ($\rho=0$) & 0.58 & -0.25 & 0.092 & 0.11 \\
\hline
\end{tabular}
\end{center}
\label{table:D_B_medium}
\end{table}%

We close this section by leaving a comment for the case of $g_{1} \neq 0$.
Let us suppose $g_{2}=0$ in this case.
Then, from Eq.~(\ref{eq:decay_width}), we obtain $(g,g_{1}/M_{\mathrm{D}},g_{2}/M_{\mathrm{D}})=(0.5,0.07,0)$ and $(0.4,0.17,0)$.
We find similarly the in-medium masses by calculating the self-energies for $\bar{\mathrm{D}}^{(\ast)}$ and $\mathrm{B}^{(\ast)}$ mesons from Eqs.~(\ref{eq:selfenergy_D}) and (\ref{eq:selfenergy_D*}).
By applying the in-medium mass formula in Eq.~(\ref{eq:mass_formula_density}), we finally get the matrix elements as summarized in Table~\ref{table:D_B_medium2}.
Interestingly, we find the same tendency that $\lambda_{1}(\rho)$ is enhanced, and $\bar{\Lambda}(\rho)$, $\lambda_{2}(\rho;m_{\mathrm{c}})$ and $\lambda_{2}(\rho;m_{\mathrm{b}})$ are reduced, as discussed above.
Thus, we expect that our conclusion will be insensitive to the small change of the parameter sets.

\begin{table}[tbp]
\caption{The matrix elements $\bar{\Lambda}(\rho)$, $\lambda_{1}(\rho)$, $\lambda_{2}(\rho,m_{\mathrm{c}})$ and $\lambda_{2}(\rho,m_{\mathrm{b}})$ with $g_{2}=0$. (Same convention as Table~\ref{table:D_B_medium}.)}
\begin{center}
\begin{tabular}{|c|c|c|c|c|}
\hline
  $(g, g_{1}/M_{\mathrm{D}},g_{2}/M_{\mathrm{D}})$ &  $\bar{\Lambda}(\rho)$ [GeV] & $\lambda_{1}(\rho)$ [GeV$^2$] & $\lambda_{2}(\rho;m_{\mathrm{c}})$ [GeV$^2$] & $\lambda_{2}(\rho;m_{\mathrm{b}})$ [GeV$^2$] \\
\hline
 $(0.5, 0.07, 0)$ & 0.50 & -0.30 & 0.056 & 0.018 \\
 $(0.4, 0.17, 0)$ & 0.52 & -0.25 & 0.059 & 0.040 \\
\hline
vacuum ($\rho=0$) & 0.58 & -0.25 & 0.092 & 0.11 \\
\hline
\end{tabular}
\end{center}
\label{table:D_B_medium2}
\end{table}%

\section{Discussion of matrix elements in heavy baryons}\label{sec:discussion}

In the previous section, 
we have discussed the medium effects in nuclear matter that $\lambda_{1}$ is enhanced and $\bar{\Lambda}$ and $\lambda_{2}(m_{\mathrm{Q}})$ are reduced, as they are compared with the values in vacuum.
We investigate whether this is generally correct, when the heavy quark exists in baryon-rich environments.
Such a tendency may be seen for systems even with a few baryon numbers.
In this section, we consider charm and bottom baryons as systems with minimal (finite) baryon number, and discuss the values of $\bar{\Lambda}$, $\lambda_{1}$ and $\lambda_{2}(m_{\mathrm{Q}})$ in them.

Let us discuss normal heavy baryons whose minimal quark configuration is Qqq.
We consider $\Lambda_{\mathrm{c}}$ and $\Lambda_{\mathrm{b}}$ baryons with $1/2^+$ as the ground states, and $\Lambda_{\mathrm{c}}^{\ast}$ and $\Lambda_{\mathrm{b}}^{\ast}$ baryons with $1/2^{-}$ and $3/2^{-}$ as the lowest excited states, in isosinglet sector.
We consider also $\Sigma_{\mathrm{c}}^{(\ast)}$ and $\Sigma_{\mathrm{b}}^{(\ast)}$ baryons with $1/2^{+}$ ($3/2^{+}$) as the ground states (the lowest excited states) in isotriplet sector \cite{Beringer:1900zz}.
We note that $\Lambda_{\mathrm{b}}^{\ast}$ baryons with $1/2^{-}$ and $3/2^{-}$ have been recently reported by LHCb as two states with masses $5911.97$ MeV and $5919.77$ MeV and their quantum numbers are not settled yet \cite{Aaij:2012da}.
However, we may naturally assign the quantum number $1/2^{-}$ ($3/2^{-}$) to the state with the smaller (larger) mass.

In isosinglet sector, 
 $\Lambda_{\mathrm{c}}$ and $\Lambda_{\mathrm{b}}$ baryons with $1/2^{+}$ are assigned to the states containing the brown muck with isospin, total angular momentum and parity $I(j^{\mathcal{P}})=0(0^{+})$.
Similarly, $\Lambda_{\mathrm{c}}^{\ast}$ and $\Lambda_{\mathrm{b}}^{\ast}$ baryons with $1/2^{-}$ and $3/2^{-}$ are assigned to the states containing the brown muck with $0(1^{-})$.
In isotriplet sector, $\Sigma_{\mathrm{c}}^{(\ast)}$ and $\Sigma_{\mathrm{b}}^{(\ast)}$ baryons with $1/2^{+}$ ($3/2^{+}$) are  assigned to the states containing the brown muck with $1(1^{+})$.
Thus, each of pairs $(\Lambda_{\mathrm{c}},\Lambda_{\mathrm{b}})$, $(\Lambda_{\mathrm{c}}^{\ast},\Lambda_{\mathrm{b}}^{\ast})$ and $(\Sigma_{\mathrm{c}}^{(\ast)},\Sigma_{\mathrm{b}}^{(\ast)})$ belongs to an independent spin multiplet.
To apply the mass formula in Eq.~(\ref{eq:mass_formula}) for each pair, 
 we use $S_{\mathrm{L}}=0$ for $(\Lambda_{\mathrm{c}},\Lambda_{\mathrm{b}})$ with $1/2^+$, and $S_{\mathrm{L}}=1$ for $(\Lambda_{\mathrm{c}}^{\ast}, \Lambda_{\mathrm{b}}^{\ast})$ and $(\Sigma_{\mathrm{c}}^{(\ast)}, \Sigma_{\mathrm{b}}^{(\ast)})$.
Consequently, we obtain the matrix elements $\bar{\Lambda}$, $\lambda_{1}$, $\lambda_{2}(m_{\mathrm{c}})$ and $\lambda_{2}(m_{\mathrm{b}})$ for each pair as shown in Table~\ref{table:charm_bottom_baryons}.
We find that $\lambda_{1}$ is larger than that of $\bar{\mathrm{D}}^{(\ast)}$ and $\mathrm{B}^{(\ast)}$ mesons, and $\lambda_{2}(m_{\mathrm{c}})$ and $\lambda_{2}(m_{\mathrm{b}})$ are smaller. 
Thus, we find that $\lambda_{1}$ in the baryons with a heavy quark is enhanced and $\lambda_{2}(m_{\mathrm{c}})$ and $\lambda_{2}(m_{\mathrm{b}})$ are reduced, in comparison with the meson cases.
This is consistent with the expectations from the analysis in nuclear matter. 
We note, however, that $\bar{\Lambda}$ in the heavy baryons are almost as twice as that in $\bar{\mathrm{D}}^{(\ast)}$ and $\mathrm{B}$ mesons.
This is simply because the number of light quarks in heavy baryons is twice of that in $\bar{\mathrm{D}}^{(\ast)}$ and $\mathrm{B}$ mesons.

Next, let us consider exotic baryons whose minimal quark configuration is $\bar{\mathrm{Q}}\mathrm{qqqq}$.
Apparently, they cannot be regarded to normal baryons with three quarks.
The states with such an exotic quark configuration can be given by the hadronic molecules composed by a $\bar{\mathrm{D}}^{(\ast)}$ ($\mathrm{B}^{(\ast)}$) meson and a nucleon $\mathrm{N}$.
In the previous works by authors and the collaborators, it was discussed that there can exist several bound/resonant states of a $\bar{\mathrm{D}}^{(\ast)}$ ($\mathrm{B}^{(\ast)}$) meson and a nucleon $\mathrm{N}$ \cite{Yasui:2009bz,Yamaguchi:2011xb,Yamaguchi:2011qw}.
Among them, let us consider the $\bar{\mathrm{D}}^{(\ast)}\mathrm{N}$ ($\mathrm{B}^{(\ast)}\mathrm{N}$)
states with $1/2^{-}$ and $3/2^{-}$ and the states with $1/2^{+}$ and $3/2^{+}$ in isosinglet sector.
Only the states with $1/2^{-}$ are bound states, and the other states are resonant states.
Because we apply the mass formula in Eq.~(\ref{eq:mass_formula}), we consider only their masses by neglecting the decay widths in the resonances.
As for the quantum numbers of the brown mucks, we assign $0(1^{+})$ for the $(1/2^{-},3/2^{-})$ states  and $0(1^{-})$ for the $(1/2^{+},3/2^{+})$ states, as discussed in Ref.~\cite{Yasui:2013vca}.
Then, 
 we obtain the matrix elements $\bar{\Lambda}$, $\lambda_{1}$, $\lambda_{2}(m_{\mathrm{c}})$ and $\lambda_{2}(m_{\mathrm{b}})$ as listed in Table.~\ref{table:charm_bottom_baryons_exotic}.
We find that $\lambda_{1}$ is enhanced, while $\lambda_{2}(m_{\mathrm{c}})$ and $\lambda_{2}(m_{\mathrm{b}})$ is reduced, both for $(1/2^{-},3/2^{-})$  and for $(1/2^{+}, 3/2^{+})$.
This is again consistent with our expectations from the results in nuclear matter.
Note that $\bar{\Lambda}$ is larger than that in $\bar{\mathrm{D}}^{(\ast)}$ and $\mathrm{B}$ mesons, because the number of light quarks in the exotic heavy baryons is larger.

A comment is in order.
In our previous works about $\bar{\mathrm{D}}^{(\ast)}\mathrm{N}$ and $\mathrm{B}^{(\ast)}\mathrm{N}$ states in Refs.~\cite{Yasui:2009bz,Yamaguchi:2011xb,Yamaguchi:2011qw}, the $1/M$-corrections were included in the kinetic energies and the mass difference between $\bar{\mathrm{D}}$ and $\bar{\mathrm{D}}^{\ast}$ ($\mathrm{B}$ and $\mathrm{B}^{\ast}$) mesons.
However, there was no inclusion of $1/M$-corrections in the interaction vertices with axial-vector current coupling to a pion, which are given in Eq.~(\ref{eq:Lagrangian_HMET}).
In this sense, the results in Refs.~\cite{Yasui:2009bz,Yamaguchi:2011xb,Yamaguchi:2011qw} does not necessarily correspond to the full $1/M$-expansions.
Nevertheless, it is interesting to observe that the $\bar{\mathrm{D}}^{(\ast)}\mathrm{N}$ and $\mathrm{B}^{(\ast)}\mathrm{N}$ states 
 cause the change of the matrix elements $\bar{\Lambda}$, $\lambda_{1}$, $\lambda_{2}(m_{\mathrm{c}})$ and $\lambda_{2}(m_{\mathrm{b}})$, as seen in the normal charm and bottom baryons.

\begin{table}[tbp]
\caption{The matrix elements $\bar{\Lambda}$, $\lambda_{1}$, $\lambda_{2}(m_{\mathrm{c}})$ and $\lambda_{2}(m_{\mathrm{b}})$ of  normal charm and bottom baryons. For comparison, the matrix elements of $\bar{\mathrm{D}}^{(\ast)}$ and $\mathrm{B}^{(\ast)}$ mesons are also shown in the last column.}
\begin{center}
\begin{tabular}{|c|c|c|c|c|}
\hline
 baryons &  $\bar{\Lambda}$ [GeV] & $\lambda_{1}$ [GeV$^2$] & $\lambda_{2}(m_{\mathrm{c}})$ [GeV$^2$] & $\lambda_{2}(m_{\mathrm{b}})$ [GeV$^2$] \\
\hline
 $\Lambda_{\mathrm{c}}$, $\Lambda_{\mathrm{b}}$ ($1/2^{+}$) & 0.88 & -0.28 & --- & --- \\
 $\Lambda^{\ast}_{\mathrm{c}}$, $\Lambda^{\ast}_{\mathrm{b}}$ ($1/2^{-}$, $3/2^{-}$) & 1.17 & -0.39 & 0.014 & 0.012 \\
 $\Sigma^{(\ast)}_{\mathrm{c}}$, $\Sigma^{(\ast)}_{\mathrm{b}}$ ($1/2^{+}$, $3/2^{+}$) & 1.09 & -0.29 & 0.028 & 0.032 \\
\hline
 $\bar{\mathrm{D}}^{(\ast)}$, $\mathrm{B}^{(\ast)}$ & 0.58 & -0.25 & 0.092 & 0.11 \\
\hline
\end{tabular}
\end{center}
\label{table:charm_bottom_baryons}
\end{table}%

\begin{table}[tbp]
\caption{The matrix elements $\bar{\Lambda}$, $\lambda_{1}$, $\lambda_{2}(m_{\mathrm{c}})$ and $\lambda_{2}(m_{\mathrm{b}})$ of  exotic charm and bottom baryons. (Same convention as Table~\ref{table:charm_bottom_baryons}.)}
\begin{center}
\begin{tabular}{|c|c|c|c|c|}
\hline
 states &  $\bar{\Lambda}$ [GeV] & $\lambda_{1}$ [GeV$^2$] & $\lambda_{2}(m_{\mathrm{c}})$ [GeV$^2$] & $\lambda_{2}(m_{\mathrm{b}})$ [GeV$^2$] \\
\hline
 $\bar{\mathrm{D}}^{(\ast)}\mathrm{N}$, $\mathrm{B}^{(\ast)}\mathrm{N}$ ($1/2^{-}$, $3/2^{-}$) & 1.48 & -0.27 & 0.050 & 0.047 \\
 $\bar{\mathrm{D}}^{(\ast)}\mathrm{N}$, $\mathrm{B}^{(\ast)}\mathrm{N}$ ($1/2^{+}$, $3/2^{+}$) & 1.50 & -0.30 & 0.053 & 0.041 \\
\hline
 $\bar{\mathrm{D}}^{(\ast)}$, $\mathrm{B}^{(\ast)}$ & 0.58 & -0.25 & 0.092 & 0.11 \\
\hline
\end{tabular}
\end{center}
\label{table:charm_bottom_baryons_exotic}
\end{table}%

\section{Summary}

We discuss the $1/M$-corrections with the heavy meson mass $M$ in the in-medium masses of $\bar{\mathrm{D}}^{(\ast)}$ ($\mathrm{B}^{(\ast)}$) mesons in nuclear matter.
Following the formalism based on the velocity-rearrangement invariance, we give the heavy meson effective Lagrangian with the $1/M$-corrections, and apply it to calculate the in-medium masses of $\bar{\mathrm{D}}^{(\ast)}$ ($\mathrm{B}^{(\ast)}$) mesons.
From the relation between the masses of heavy hadrons and the gluon dynamics,
we obtain the modifications of the gluon fields in nuclear matter.
We show that the effect of scale anomaly becomes suppressed in nuclear matter,
and  also show that the contribution to the in-medium mass from the chromoelectric gluon is enhanced, while that from the chromomagnetic gluon is reduced.
We discuss the cases of the heavy baryons with a heavy quark both in normal sector and in exotic sector.

As we have emphasized, the mass formula in Eq.~(\ref{eq:mass_formula}) holds for any states with a heavy quark in various environments.
In the present study, we have investigated the $\bar{\mathrm{D}}^{(\ast)}$ and $\mathrm{B}^{(\ast)}$ mesons in nuclear matter at zero temperature, and have discussed the effects of scale anomaly and chromoelectric and chromomagnetic gluons. 
Similar discussions will be applied to other charm and bottom hadrons such as $\Lambda_{\mathrm{c}}$ and $\Lambda_{\mathrm{b}}$ baryons as well as $\mathrm{D}^{(\ast)}$ and $\bar{\mathrm{B}}^{(\ast)}$ mesons in nuclear medium, when appropriate dynamical processes are considered.
Moreover, the present analysis can be applied also to the heavy quark systems in the deconfinement phases with finite temperature and/or finite density.
Then, we will be able to discuss the scale anomaly and the chromoelectric and chromomagnetic gluons in various phases.
Those studies will be interesting in experiments in hadron reactions at J-PARC and GSI-FAIR and also in heavy ion collisions at RHIC and LHC \cite{Cho:2010db,Cho:2011ew}.

\begin{acknowledgments}
This work is supported in part by Grant-in-Aid for Scientific Research on 
Priority Areas ``Elucidation of New Hadrons with a Variety of Flavors 
(E01: 21105006)" (S.~Y.) and by Grant-in-Aid for Scientific Research from MEXT
(Grants No. 22740174 (K.~S.)).
\end{acknowledgments}


\end{document}